\documentclass{ws-ijmpe}

\begin{document}
\markboth{J.-P. Chen}
{Moments of Spin Structure Functions: Sum Rules and Polarizabilities}
%
%
\catchline{}{}{}{}{}
%
%
\title{Moments of Spin Structure Functions: Sum Rules and Polarizabilities}
\author{J.-P. Chen}
\address{Thomas Jefferson National Accelerator Facility\\
12000 Jefferson Avenue, Newport News, Virginia 23606, USA
\\
jpchen@jlab.org}
\maketitle
\begin{history}
\received{received date}
\revised{revised date}
\end{history}
\begin{abstract}
Nucleon structure study is one of the most important research areas 
in modern physics and has challenged us for decades. 
Spin has played an essential role and often brought surprises and puzzles 
to the investigation of the nucleon structure and the strong interaction.
New experimental data on nucleon spin structure at low 
to intermediate momentum
transfers combined with existing high momentum transfer data offer a 
comprehensive picture in the strong region of the interaction and 
of the transition region from the strong to the
asymptotic-free region. Insight for some aspects of the theory for the strong interaction, Quantum Chromodynamics (QCD), is gained by
exploring lower moments of spin structure functions and  their corresponding sum rules (i.e.
the Bjorken, Burkhardt-Cottingham, Gerasimov-Drell-Hearn (GDH), and the 
generalized GDH). These moments are expressed in
terms of an operator-product expansion using quark and gluon degrees of
freedom at moderately large momentum transfers.
The higher-twist contributions have been examined through
the evolution of these moments as the momentum transfer varies from higher to lower values.
Furthermore, QCD-inspired low-energy effective theories, which explicitly include chiral 
symmetry breaking, are tested at low momentum transfers.
The validity of these theories is further examined as  the momentum transfer 
increases to
moderate values. It is found that chiral perturbation theory calculations 
agree reasonably well 
with the first moment of the spin structure function $g_1$   
at low momentum transfer of 0.05 - 0.1 GeV$^2$ but fail to reproduce some of 
the higher moments, noticeably, 
the neutron data in the case 
of the generalized polarizability $\delta_{LT}$. The Burkhardt-Cottingham sum 
rule has been verified with good accuracy in a wide range of $Q^2$ 
assuming that no singular behavior of the structure functions 
is present at very high excitation energies. 
\end{abstract}

\section{Introduction}	
The nucleons (proton and neutron), the basic building blocks of nuclear 
matter, account for more than $99.9\%$ of the mass of the visible matter in 
the universe.
Understanding the nucleon structure is one of the most important issues 
in modern physics and has challenged us for decades. The internal structure of
the nucleon
presents a wealth of fundamental questions, that have a deep impact on our 
understanding of Nature and the universe we live in. 

The strong interaction,
that is mostly responsible for the structure of the nucleon, poses a special
challenge to us due to the strong and non-linear nature of the force. 
Quantum Chromodynamics 
(QCD) is the accepted theory for the strong interaction. 
Quarks and gluons are the fundamental particles and force 
mediators, while nucleons and mesons (pions, kaons, etc.), collectively 
called hadrons, are composite particles and effective force mediators.  
QCD has been well tested at high energy (asymptotic-free) when 
the interaction is relatively weak due to the running of strong coupling 
constant. 
However, at low energy (scale of the size of a hadron),
the strong force becomes truly strong, leading to chiral symmetry breaking
and the confinement of the quarks and gluons inside the hadron where 
none can escape to become free particles. In the strong region 
of QCD, the usual calculation technique of perturbative expansion fails, and 
QCD can not be easily solved. Effective field theories, such as chiral 
perturbation theory~\cite{CPT}, are often used. Recently, great progress has been made in
the advancement of lattice QCD~\cite{LQCD}, which has begun to provide some 
calculations 
of QCD in the strong region. 
Another approach~\cite{SD} is to stay with the continuous field theory
by using the Schwinger-Dyson equations. 
Newly developed anti-de Sitter/conformal field theory (AdS/CFT)~\cite{AdS} 
might provide another useful tool to study strong QCD. 
  
In addition to the truly strong region, understanding the 
transition from weak (perturbative) to strong (non-perturbative) region 
is also very important. One approach is to start from high energy where the
quarks and gluons (collectively called partons) are asymptotically free. 
With the help of the Operator Product Expansion (OPE)~\cite{SV,stein,Ji93}, 
one can 
systematically expand to a twist series, where the leading twist term gives 
the parton distributions and the higher-twist terms characterize the 
quark-gluon and 
quark-quark correlations. Lattice QCD can also provide useful calculations.
Constituent-quark models are also often used to provide intuitive 
pictures in this region. 

Experimentally, deep-inelastic lepton scattering 
(DIS) together with other
processes, such as Drell-Yan, direct photon and heavy quark productions, have
been powerful tools to study nucleon structure. The 
unpolarized structure functions extracted from the last five decades, covering 
five-order of magnitudes in both $Q^2-$ and $x-$ranges, are the most extensive
set of precision data in nucleon structure study. The longitudinally polarized
structure functions extracted from the last three decades have reached reasonable
precision, covering two to three-order of magnitudes in $Q^2-$ and $x-$ ranges. Transversely polarized measurements have been the recent focus, but 
are still scarce. The lower moments of the structure functions provide the most
direct tests and comparisons with theoretical calculations through sum rules.

In the study of nucleon structure, spin has played an essential role and often
brought big surprises.  
Spin was first introduced as an internal property of a particle, related to 
the magnetic moment, to explain the famous Stern-Gerlach results~\cite{SG} 
that the 
silver atoms split into two beams after passing through an inhomogeneous 
magnetic field. The discovery of the anomalous magnetic moment of the 
proton~\cite{AMM} was the first big spin surprise. This discovery was the 
beginning of the (indirect) study of nucleon structure, since the anomalous 
magnetic moment was one piece of evidence for the internal structure of the 
nucleon. 
Later on, it was
related to the integration of the excitation spectrum of nucleon spin 
structure by the sum rule of Gerasimov, Drell and Hearn 
(GDH)~\cite{GDH,GDH2}. The direct study of nucleon structure 
started decades
later when Hofstadter~\cite{Hof} used elastic electron scattering to 
directly measure the form factors of the proton. 

In the last thirty years, the spin structure of the nucleon has led to very 
productive experimental and theoretical activity with exciting results and 
new challenges\cite{KCL09}. This
investigation has included a variety of aspects such as
testing QCD in its perturbative region 
{\emph {via}} spin sum rules (like the Bjorken sum rule\cite{bjo66}) and understanding how the spin of the nucleon is
built from the intrinsic degrees of freedom of the theory: quarks and gluons. 
Recently, results from a new generation of experiments performed at
Jefferson Lab seeking to probe the theory in its non-perturbative 
and transition regions
 have reached a mature state.  The low momentum-transfer results offer insight in 
a region characterized by the collective behavior of the nucleon constituents and their interactions.
In this region, it has been more economical to describe the nucleon using effective degrees
of freedom like mesons and  constituent quarks rather than current quarks and gluons. 
Furthermore, 
distinct features seen in the nucleon response to the electromagnetic probe, depending on the 
resolution  of the probe, point clearly to different 
regions of description, {\it {i.e.}} a scaling region where quark-gluon correlations are suppressed 
versus a coherent region where long-range interactions give rise to the 
static properties of the nucleon. 

In this review we describe an investigation~\cite{e97110}$^{-}$\cite{BJ}
 of the spin structure of the nucleon through  
measurements of the helicity-dependent photoabsorption cross sections or asymmetries by using virtual photons  
covering a wide resolution spectrum. These observables are used to extract the
spin structure 
functions $g_1$ and $g_2$ and evaluate their moments. These moments are key 
ingredients 
to test QCD sum rules and unravel some aspects of the quark-gluon structure of the nucleon. 

\section{Formalism}

\begin{figure}[htb!]
\begin{center}
\begin{minipage}[t]{9 cm}
\epsfig{file=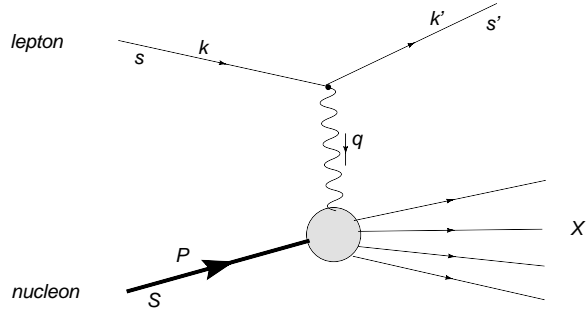,scale=0.7}
\end{minipage} 
\begin{minipage}[t]{16.5 cm}
\caption{Feynman diagram for deep-inelastic lepton-hadron
scattering} \label{fig1}
\end{minipage}
\end{center}
\end{figure}

Leptons do not involve the strong interaction directly. With the
electromagnetic 
interaction well-understood, lepton beams have been a very powerful precision
probe to study nucleon structure. 
Consider deep-inelastic scattering of polarized leptons on
polarized nucleons. We denote by $m$ the lepton mass,  $
k~(k^\prime$) the initial (final) lepton four-momentum and
$s~(s^\prime)$ its covariant spin four-vector, such that $s \cdot
k$ = 0 $(s^\prime \cdot  k^\prime = 0)$ and $s \cdot s = - 1$
$(s^\prime \cdot s^\prime = -1)$; the nucleon mass is $M$ and the
nucleon four-momentum and spin four-vector are, respectively, $P$
and $S$. Assuming, as is usually done, one-photon exchange
 (see Fig.~\ref{fig1}), the
differential cross section for detecting the final polarized
lepton within the solid angle $d\Omega$ and the final energy range
$(E^\prime,~E^\prime + dE^\prime)$ in the laboratory frame $P =
(M, \bm{0}), ~k = (E, \bm{k}), ~k^\prime = (E^\prime,\bm{k}^\prime)$, 
can be written as

\begin{equation} \label{eq:DIS}
\frac{d^2 \sigma}{ d \Omega dE'} = \frac{\alpha^2}{ 2 M q^4}
\frac{E'}{E} L_{\mu \nu} W^{\mu \nu},
\end{equation}
where $q =  k- k'$ and $\alpha$ is the fine structure constant.

The leptonic tensor $L_{\mu\nu}$ is given by
\begin{equation}\label{eq:L-tensor} L_{\mu\nu} ( k, s; k^\prime)=
\sum_{s'} \,[\bar{u} (k^\prime, s^\prime)
~\gamma_{\mu}~u(k,s)]^\ast ~[\bar{u}(k^\prime,
s^\prime)~\gamma_{\nu}~u(k,s)]  \end{equation}
 and can be split into
symmetric $(S)$ and antisymmetric $(A)$ parts under $\mu$ and $\nu$
interchange: 
\begin{equation} \label{eq:L-split}
 L_{\mu\nu}(k, s; k^\prime)=  2 \{
L^{(S)}_{\mu\nu}~(k;k^\prime) + iL^{(A)}_{\mu\nu}~(k,s;k^\prime)\}
 \end{equation}
where $u(k,s)$ and $\bar{u}(k^\prime,s^\prime)$ are electron spinors, and
\begin{eqnarray} \label{eq:L-S}
L^{(S)}_{\mu\nu}~(k;k^\prime) &=& k_\mu k^\prime_\nu +
k^\prime_\mu k_\nu -
g_{\mu\nu}~(k\cdot k^\prime - m^2)  \\
\label{eq:L-A}
L^{(A)}_{\mu\nu}~(k,s;k^\prime) &=&
m~\varepsilon_{\mu\nu\alpha\beta}~s^\alpha~q^{\beta} .
\end{eqnarray}
The unknown hadronic tensor $W_{\mu\nu}$ describes the interaction
between the virtual photon and the nucleon and depends upon four
scalar structure functions: the unpolarized  functions $F_{1,2}$
and the spin-dependent  functions $ g_{1,2}$ (ignoring parity-violating 
interactions). These functions contain 
information on the internal structure of the nucleon. They can
be experimentally measured and then be studied in theoretical models, 
such as the QCD-parton model. The functions can only depend on 
the scalars $q^2$ and $ q\cdot P$. Usually people work with
\begin{equation} \label{eq:def-Q,xBj}
  Q^2 \equiv -q^2 \quad \textrm{and} \quad x \equiv Q^2/2q \cdot
  P= Q^2/2M\nu
\end{equation}
where $\nu=E-E'$ is the energy of the virtual photon in the Lab
frame. The variable $x$, known as ``$x$-Bjorken", is the fraction of momentum
carried by the struck quark in the simple parton model.  
We also refer to the invariant mass of the (unobserved)
final state, $W = (P+q)^2 = M^2 + 2M\nu - Q^2$.

Analogous to Eq.~(\ref{eq:L-split}) one has  

\begin{equation}
\label{eq:Wmunu} W_{\mu\nu}(q; P,S) = W^{(S)}_{\mu\nu}(q;P) +
i~W_{\mu\nu}^{(A)}(q; P, S) .  
\end{equation}

The symmetric part, relevant to unpolarized DIS, is given by
\begin{eqnarray} \label{eq:WS}
 W^{(S)}_{\mu\nu}(q;P) &=&
  2 \left[ \frac{q_\mu q_\nu}{q^2} - g_{\mu \nu}
\right]\,F_1(x,Q^2) \\ \nonumber
&& + \frac{2}{M \nu} \left[P_\mu -\frac{P\centerdot q}{q^2}q_\mu
\right]\left[P_\nu -\frac{P\centerdot q}{q^2}q_\nu
\right]\,F_2(x,Q^2) . \end{eqnarray}
The antisymmetric part, relevant for polarized DIS, can be written as
\begin{equation} \label{eq:WAG12}
 W^{(A)}_{\mu\nu }(q; P, s) = 2~\varepsilon_{\mu\nu\alpha\beta}~ q^\alpha \Biggl\{ M^2 S^\beta G_1
(\nu, Q^2) +   \left[M \nu S^\beta - (S\cdot q) \, P^\beta \right] G_2 (\nu, Q^2)\Biggl\} \end{equation}
or, in terms of the scaling functions $g_{1,2}$
\begin{equation} \label{eq:g12G12}
g_1(x,Q^2)=M^2 \nu G_1(\nu,Q^2), \qquad g_2(x,Q^2)= M \nu^2 G_2(\nu,Q^2) , 
\end{equation}
\begin{equation} \label{eq:WA}
 W^{(A)}_{\mu\nu }(q; P, s) = \frac{2M}{ P\cdot
q}~\varepsilon_{\mu\nu\alpha\beta}~ q^\alpha\Biggl\{ S^\beta g_1
(x, Q^2) +   \left[S^\beta - \frac{(S\cdot q) \, P^\beta}{
(P\cdot q)}\right] g_2 (x, Q^2)\Biggl\}\,. \end{equation}

Note that these expressions are electromagnetically gauge-invariant:
\begin{equation} q^\mu W_{\mu\nu} =0 . \end{equation}
In the Bjorken limit, or deep-inelastic regime,
\begin{equation}
\label{eq:Bjlim} Q^2 \to \infty\quad , \quad \nu \to \infty \quad , \quad x \, \, \rm {fixed}, \end{equation}
the structure functions $F_{1,2}$ and $g_{1,2}$
 are known to approximately scale, i.e., vary very slowly with $Q^2$
at fixed $x$ -- in the simple parton model they scale exactly; in
QCD their $Q^2$-evolution can be calculated perturbatively.

The differential cross section for unpolarized leptons scattering off
an unpolarized target can be written as 
\begin{equation} \label{eq:Xsec-unpol}
\frac{d^2\sigma}{d\Omega~dE^\prime}
 (k, s, P, - S; k^\prime)  =
\frac{\alpha^2}{2Mq^4}~\frac{E'}{
E}~2L_{\mu\nu}^{(S)}~W^{\mu\nu(S)}.
\end{equation}
While differences of cross sections with opposite target spins are given
by
\begin{equation} \label{eq:Xsec-diff}
\left[\frac{d^2\sigma}{d\Omega~dE^\prime}
 (k, s, P, - S; k^\prime) -\frac{d^2\sigma}{d\Omega
~dE^\prime}~(k, s, P, S; k^\prime)\right] =
\frac{\alpha^2}{2Mq^4}~\frac{E'}{
E}~4L_{\mu\nu}^{(A)}~W^{\mu\nu(A)}.
\end{equation}

After some algebra~\cite{Anselmino:1994gn},
one obtains from Eqs.~(\ref{eq:L-S},\ref{eq:L-A},\ref{eq:WS},\ref{eq:WA},\ref{eq:Xsec-unpol},\ref{eq:Xsec-diff}) 
expressions for
the unpolarized cross section and polarized cross-section 
differences (note that the lepton-mass terms are neglected):
 \begin{itemize}
 \item The unpolarized cross section is given by
 \begin{equation}
\label{eq:UnpolXsec}
 \frac{d^2 \sigma_{unpold}}{d\Omega~dE^\prime}=\Bigl(\frac{d\sigma}{d\Omega}\Bigr)_{Mott}
\Bigl[
 \frac {2}{M}F_1(x,Q^2)\tan^2 \frac{2}{\theta}+\frac{1}{\nu}F_2(x,Q^2)
\Bigr]
 \end{equation}
where $\theta$ is the scattering angle in the 
laboratory frame and the Mott cross section is given by
 \begin{equation}
\Bigl(\frac{d\sigma}{d\Omega}\Bigr)_{Mott}=\frac{\alpha^2\cos^2\frac{\theta}{2}}{4E^2\sin^4\frac{\theta}{2}}.
 \end{equation}

 \item For the lepton and  target
nucleon polarized longitudinally, i.e. along or opposite to the
direction of the lepton beam,  the
cross-section difference under reversal of the
nucleon's spin direction (indicated by the double arrow) is given by
\begin{equation} \label{eq:LongXsec}
\frac{d^2\sigma^{\begin{array}{c}\hspace*{-0.2cm}\to\vspace*{-0.2cm}\\
\hspace*{-0.2cm}\Leftarrow\end{array}}}{d\Omega~dE^\prime}-
\frac{d^2\sigma^{\begin{array}{c}\hspace*{-0.2cm}\to\vspace*{-0.2cm}\\
\hspace*{-0.2cm}\Rightarrow\end{array}}}{d\Omega~dE^\prime}
 =
-
\frac{4\alpha^2E^\prime}{\nu EQ^2}\Biggl[(E+E^\prime \cos \theta)~g_1(x,Q^2) 
- 2Mxg_2(x,Q^2) \Biggl] ~.
\end{equation}
\item For nucleons polarized transversely in the scattering plane,
 one finds
\begin{equation} \label{TransXsec}
\frac{d^2 \sigma^{\to\Uparrow}}{d\Omega~dE^\prime} - \frac{d^2
\sigma^{\to\Downarrow}}{d\Omega~dE^\prime} = 
- \frac{4
\alpha^2 {E^\prime}^2}{\nu EQ^2}~\left[g_1(x,Q^2) + \frac{2ME}{\nu}g_2(x,Q^2)\right]\,.
\end{equation}
\end{itemize}


These two independent observables allow measurement
of both $g_1 $ and $g_2$
(as has been done at SLAC and in Jefferson Lab's Halls A and C),
but the transverse cross section difference is
generally smaller
because of kinematic factors and therefore more
difficult to measure. Only in the past decade has it been
possible to gather precise information on $g_2$, which indeed turns out to be
usually smaller than $g_1$ in the deep-inelastic region.

Experimental results are often presented in the form of
asymmetries, which are ratios of the cross-section differences to
the unpolarized cross section.

For a longitudinally polarized target, the measured asymmetry is
\begin{equation} \label{eq:Apar} A_\| \equiv
\frac{d\sigma^{\begin{array}{c}\hspace*{-0.2cm}\to\vspace*{-0.2cm}\\
\hspace*{-0.2cm}\Leftarrow\end{array}}-
d\sigma^{\begin{array}{c}\hspace*{-0.2cm}\to\vspace*{-0.2cm}\\
\hspace*{-0.2cm}\Rightarrow\end{array}}}{2\,d\sigma_{unpold}}
\end{equation}
and for a transversely polarized target
\begin{equation} \label{eq:Atrans} A_{\perp}\equiv
\frac{d\sigma^{\to\Uparrow} - d
\sigma^{\to\Downarrow}}{2d\sigma_{unpold}} . \end{equation}

It is customary to introduce the (virtual) photon-nucleon asymmetries
$A_{1,2}$:
\begin{equation} \label{eq:Aigi} A_1= \frac{\sigma_{TT}}{\sigma_T}
=\frac{g_1 - \gamma ^2 g_2}{F_1} 
\end{equation}
 and
 \begin{equation} \label{eq:A_2gi}
 A_2= \frac{\sigma_{LT}}{\sigma_T}
=\gamma \, \big[\frac{g_1 + g_2}{F_1}\big], 
 \end{equation}
where 
 \begin{equation} \label{eq:gamma}
     \gamma    = \frac{2Mx}{Q}= \frac{Q}{\nu}~,
\end{equation}
and $\sigma_T$, $\sigma_{TT}$ and $\sigma_{LT}$ are the virtual photon cross sections
(see below for the definitions of virtual photon cross sections).

 From Eqs.~(\ref{eq:UnpolXsec},\ref{eq:LongXsec},\ref{TransXsec}), it follows that
\begin{equation} \label{eq:AlongA12}
 A_\|= D\,(A_1 + \eta A_2) \end{equation}
 and
  \begin{equation} \label{eq:AperpA12}
A_{\perp}=d(A_2 - \xi A_1) \end{equation}
where
\begin{equation} \label{Dnew}
D= \frac{y[(1 + \gamma^2
y/2)(2-y)-2y^2m^2/Q^2]}{y^2(1-2m^2/Q^2)(1+\gamma^2) + 2
(1+R)(1-y-\gamma^2 y^2/4)} 
\end{equation}
\begin{equation} \label{dnew}
d= \left[\frac{[1+\gamma^2y/2(1 + 2m^2y/Q^2)]
\sqrt{1-y-\gamma^2y^2/4}}{(1-y/2)(1+\gamma^2y/2) -
y^2m^2/Q^2}\right]\,D
\end{equation}
\begin{equation} \label{etanew}
\eta=\gamma \,\frac{ [1-y-y^2(\gamma^2/4 + m^2/Q^2)
]}{(1-y/2)(1+\gamma^2y/2) - y^2m^2/Q^2}
\end{equation}
\begin{equation} \label{xinew}
\xi =\gamma \frac{1-y/2 -y^2m^2/Q^2}{1 + \gamma^2y/2(1
+2m^2y/Q^2)} ,
\end{equation}
where
\begin{equation}
y=\frac{\nu}{E},
\end{equation}
and 
\begin{equation}
R=(1+\gamma^2)\left(\frac{F_2}{2xF_1}\right)-1=\frac{\sigma_L}{\sigma_T}
\end{equation}
is the ratio of the longitudinal and transverse virtual photon cross sections.

There exists a restrictive bound on $A_2$~\cite{Soffer:1999zv,Artru:2008cp}:
 \begin{equation} \label{eq:A2bound}
 |A_2|\leq \sqrt{R\,(1+A_1)/2} . \end{equation}

In the virtual photon notation, the inclusive inelastic cross section
can be written in terms of a virtual photon flux factor and four partial cross
sections\cite{DW08}

 \begin{equation}
\label{eq:VphotonXsec}
 \frac{d^2 \sigma}{d\Omega~dE^\prime}=
\Gamma_V \sigma(\nu,Q^2),
 \end{equation}

 \begin{equation}
\label{eq:4Xsec}
 \sigma = \sigma_T + \epsilon \sigma_L 
- hP_x \sqrt{2\epsilon(1-\epsilon)} \sigma_{LT} 
- hP_z \sqrt{1-\epsilon^2} \sigma_{TT},
 \end{equation}
with the photon polarization
 \begin{equation}
\label{eq:epsilon}
\epsilon=\frac{1}{1+2(1+\nu^2/Q^2)\tan^2 (\theta/2)}~,
 \end{equation}
and the flux factor
 \begin{equation}
\label{eq:flux}
\Gamma_V=\frac{\alpha E^\prime K}{2\pi^2 E Q^2 (1-\epsilon)}~,
 \end{equation}
where $K$ is the ``equivalent photon energy'' and here we will use the 
definition according to Hand\cite{hand}
 \begin{equation}
\label{eq:Kfactor}
K=\nu(1-x)=\frac{W^2-M^2}{2M}~.
 \end{equation}
$h=\pm 1$ refers to the two helicity states of the (relativistic) lepton.
$P_z$ is the target polarization along  
the direction of the virtual photon and $P_x$, perpendicular to that direction 
in the scattering plane of the electron. 

The partial cross sections are related to the structure functions as follows:
 \begin{equation}
\label{eq:sigmaT}
\sigma_T=\sigma_{1/2}+\sigma_{3/2}=\frac{4 \pi^2 \alpha}{MK} F_1~,
 \end{equation}
 \begin{equation}
\label{eq:sigmaL}
\sigma_L=\frac{4 \pi^2 \alpha}{MK}
\bigl [\frac{1+\gamma^2}{\gamma^2 \nu} F_2 - \frac{1}{M} F_1 \bigr ]~,
 \end{equation}
 \begin{equation}
\label{eq:sigmaTT}
\sigma_{TT}=\sigma_{1/2}-\sigma_{3/2}=\frac{4 \pi^2 \alpha}{MK}
(g_1-\gamma^2 g_2)~,
 \end{equation}
 \begin{equation}
\label{eq:sigmaLT}
\sigma_{LT}=\frac{4 \pi^2 \alpha}{MK}
\gamma (g_1+g_2)~,
 \end{equation}
where $\sigma_{1/2}$ and $\sigma_{3/2}$ are the helicity cross sections
with 1/2 and 3/2 referring to the total helicity of the (virtual) 
photon-nucleon system.


Note that in the above we have kept terms of order $M^2/Q^2$ and
smaller. They are sometimes necessary in order to extract the
correct experimental values of the structure functions from the
measured asymmetries. However, the QCD analysis of the
structure functions is often carried out at \emph{leading twist} only,
ignoring higher-twist terms, i.e., higher order in $M^2/Q^2$. This is clearly
inconsistent in cases where the above terms could be important, 
for example Jefferson Lab experiments.

\section{Sum rules and Moments}

Sum rules involving the spin structure of the nucleon offer an important opportunity to study QCD. In recent years
the Bjorken sum rule at large $Q^2$ and 
the GDH sum rule at $Q^2=0$
have attracted large experimental\cite{ahr01,dutz03,dutz04} and theoretical\cite{dre01} efforts that have provided us with rich information. 
This first type of sum rules relates the moments of
the spin structure functions (or, equivalently, the spin-dependent
virtual photon cross sections) to the nucleon's static properties.
The second type of sum rules, such as the generalized GDH sum rule\cite{ji01,ans89} or
the polarizability sum rules\cite{dre03,dre04}, relate
the moments of the spin structure functions to real or virtual 
Compton amplitudes, which can be calculated theoretically. 
Both types of sum rules are based on ``unsubtracted'' dispersion relations
and the optical theorem\cite{BD}. 
The first type of sum rules uses one more general 
assumption
such as a low-energy theorem\cite{low} for the 
GDH sum rule and the Operator Production Expansion
(OPE)\cite{ope} for the Bjorken sum rule to relate the Compton amplitude to a 
static property.

The formulation below follows closely Refs. \cite{dre03,dre04}. 
Consider the forward doubly-virtual Compton scattering (VVCS)
of a virtual photon with space-like four-momentum q, i.e., 
$q^2=\nu^2-{\vec q}^2 = -Q^2 < 0$.
The absorption of a virtual photon on
a nucleon is related to inclusive electron scattering. As discussed in the previous section, the inclusive
cross section contains four partial cross sections (or structure functions):
$\sigma_T$, $\sigma_L$, $\sigma_{TT}$, $\sigma_{LT}$, (or $F_1$, $F_2$, $g_1$,
$g_2$).
In this review, we will concentrate on the spin-dependent functions, 
$\sigma_{TT}$, $\sigma_{LT}$ (or $g_1$, $g_2$).
In the following discussion, we will start with the
general situation, i.e., sum rules valid for all $Q^2$, then discuss the two
limiting cases at low $Q^2$ and at high $Q^2$.

Considering the spin-flip VVCS amplitude $g_{TT}$ and assuming it has an appropriate 
convergence behavior at high energy,  
an unsubtracted dispersion relation leads to the following equation for
$g_{TT}$:
\begin{equation}
{\rm Re}[{g}_{TT}(\nu,Q^2)-g^{pole}_{TT}(\nu,Q^2)]
=
(\frac{\nu}{2 \pi^2}){\cal P}\int^{\infty}_{\nu_0}\frac{K(\nu',Q^2)
\sigma_{TT}(\nu',Q^2)}{\nu'^2-\nu^2}d\nu', 
\label{eq:gTT}
\end{equation}
where $g_{TT}^{pole}$ is the nucleon pole (elastic) contribution, ${\cal P}$ 
denotes the principal value integral.
The lower limit of the integration $\nu_0$ is the pion-production threshold on
the nucleon.
A low-energy expansion gives:
\begin{equation}
{\rm Re}[g_{TT}(\nu,Q^2)-g^{pole}_{TT}(\nu,Q^2)]=
(\frac{2\alpha}{M^2})I_{TT}(Q^2)\nu+\gamma_{TT}(Q^2)\nu^3+O(\nu^5).
\label{eq:gTT2}
\end{equation}
$I_{TT}(Q^2)$ is the coefficient of the $O(\nu)$ term of
the Compton amplitude.
Eq.~(\ref{eq:gTT2}) defines the generalized forward spin 
polarizability $\gamma_{TT}(Q^2)$ (or $\gamma_0(Q^2)$ as it was used in Refs.
\cite{e94010-3,dre03}).
Combining Eqs.~(\ref{eq:gTT}) and (\ref{eq:gTT2}), the $O(\nu)$ term yields a sum rule 
for the generalized GDH integral\cite{dre01,ji01}:
\begin{eqnarray}
I_{TT}(Q^2) 
&=& \frac{M^2}{4 \pi^2 \alpha} \int_{\nu_0}^{\infty}\frac{K(\nu,Q^2)}{\nu}
\frac{\sigma_{TT}} 
{\nu} d\nu \nonumber \\
 =&& \hspace{-0.5cm} \frac{2M^2}{Q^2} \int_0^{x_0} \Bigr [g_1(x,Q^2) - 
\frac{4M^2}{Q^2} 
x^2 g_2(x,Q^2)\Bigl ] dx.
\label{eq:gdhsum_def1}
\end{eqnarray} 
As $Q^2 \rightarrow 0$, the low-energy theorem relates I(0) to the anomalous magnetic moment of the nucleon, $\kappa$, and 
Eq.~(\ref{eq:gdhsum_def1}) 
becomes the
original GDH sum rule\cite{GDH,GDH2}:
\begin{equation}
I(0) =\int_{\nu_0}^{\infty}\frac{\sigma_{1/2}(\nu)-\sigma_{3/2}(\nu)}{\nu}
d\nu
 = -\frac{2 \pi^2 \alpha \kappa^2 }{M^2}.
\label{eq:gdh}
\end{equation}
The $O(\nu^3)$ term yields a sum 
rule for the generalized forward spin polarizability\cite{dre03,dre04}:
\begin{eqnarray}
\gamma_{TT}(Q^2)&=&
(\frac{1}{2\pi^2})\int^{\infty}_{\nu_0}\frac{K(\nu,Q^2)}{\nu}
\frac{\sigma_{TT}(\nu,Q^2)}{\nu^3}d\nu \nonumber \\
=&&\hspace{-0.5cm}\frac{16 \alpha M^2}{Q^6}\int^{x_0}_0 x^2 \Bigl [g_1(x,Q^2)-\frac{4M^2}{Q^2}
x^2g_2(x,Q^2)\Bigr ] dx. 
\end{eqnarray} 

Considering the longitudinal-transverse interference amplitude $g_{LT}$,
with the same assumptions, one obtains:
\begin{equation}
{\rm Re}[g_{LT}(\nu,Q^2)-g^{pole}_{LT}(\nu,Q^2)]=
(\frac{2\alpha}{M^2})Q I_{LT}(Q^2)+Q \delta_{LT}(Q^2)\nu^2+O(\nu^4) 
\end{equation}
where the $O(1)$ term leads to a sum rule for $I_{LT}(Q^2)$, which is related 
to the $\sigma_{LT}$ integral over the excitation spectrum:
\begin{eqnarray}
I_{LT}(Q^2)&=&\frac{M^2}{4\pi^2 \alpha}
\int^{\infty}_{\nu_0}\frac{K(\nu,Q^2)}{\nu}
\frac{\sigma_{LT}(\nu,Q^2)}{Q}d\nu \nonumber \\ 
= && \hspace{-0.5cm}\frac{2M^2}{Q^2}\int^{x_0}_0 x^2 \Bigl [g_1(x,Q^2)+g_2(x,Q^2)
\Bigr ] dx.
\end{eqnarray}
The $O(\nu^2)$ term leads to the generalized longitudinal-transverse
polarizability\cite{dre03,dre04}:
\begin{eqnarray}
\delta_{LT}(Q^2)&=&
(\frac{1}{2\pi^2})\int^{\infty}_{\nu_0}\frac{K(\nu,Q^2)}{\nu}
\frac{\sigma_{LT}(\nu,Q^2)}{Q \nu^2}d\nu \nonumber \\
=&&\hspace{-0.5cm}\frac{16 \alpha M^2}{Q^6}\int^{x_0}_0 x^2 \Bigl [g_1(x,Q^2)+g_2(x,Q^2)
\Bigr ] dx.   
\end{eqnarray}

Alternatively, we can consider the covariant spin-dependent VVCS amplitudes 
$S_1$ and $S_2$, which are related to the spin-flip amplitudes 
$g_{TT}$ and $g_{LT}$:
\begin{equation}
S_1(\nu,Q^2)=
\frac{\nu M}{\nu^2+Q^2} \Bigr[g_{TT}(\nu,Q^2)+\frac{Q}{\nu}g_{LT}(\nu,Q^2)
\Bigl ], \nonumber
\end{equation}
\begin{equation}
S_2(\nu,Q^2)=
-\frac{M^2}{\nu^2+Q^2} \Bigr[g_{TT}(\nu,Q^2)-\frac{\nu}{Q}g_{LT}(\nu,Q^2)
\Bigl ].
\end{equation}
The dispersion relation with the same assumptions leads to
\begin{equation}
{\rm Re}[S_1(\nu,Q^2)-S^{pole}_1(\nu,Q^2)] =
\frac{4\alpha}{M}I_1(Q^2)+ \gamma_{g_1}(Q^2)\nu^2
+O(\nu^4), 
\end{equation}
where the $O(1)$ term leads to a sum rule for $I_1(Q^2)$:
\begin{equation}
I_1(Q^2)= 
\frac{2M^2}{Q^2}\int_0^{x_0}g_1(x,Q^2)dx.
\end{equation}
The $O(\nu^2)$ term leads to the generalized $g_1$
polarizability:
\begin{eqnarray}
\gamma_{g_1}(Q^2)&=&
\frac{16\pi \alpha M^2}{Q^6}
\int^{x_0}_0 x^2 g_1(x,Q^2) dx   \nonumber \\
&=& \delta_{LT} + \frac{2\alpha}{M^2 Q^2}\Bigr (I_{TT}(Q^2)-I_1(Q^2)\Bigl ).
\end{eqnarray}

For $S_2$, assuming a Regge behavior at $\nu \rightarrow \infty$ given by
$S_2 \rightarrow \nu^{\alpha_2}$ with $\alpha_2 < -1$, the unsubtracted
dispersion relations for $S_2$ and $\nu S_2$, without the elastic pole 
subtraction, lead to a ``super-convergence
relation'' that is valid for any value of $Q^2$:
\begin{equation}
\label{eq:bc}
\int_0^{1}g_2(x,Q^2)dx=0,
\end{equation}
which is the Burkhardt-Cottingham (BC) sum rule\cite{BC}. This expression
can also be written as
\begin{equation}
I_2(Q^2)=\frac{2M^2}{Q^2}\int_0^{x_0}g_2(x,Q^2)dx=\frac{1}{4}F_P(Q^2)
\Bigl (F_D(Q^2)+F_P(Q^2)\Bigr ),
\label{eq:bc_noel}
\end{equation}
where $F_P$ and $F_D$ are the Pauli and Dirac form factors for elastic
e-N scattering.  

The low-energy expansion of the dispersion relation leads to
\begin{eqnarray}
&{\rm Re}&\bigl [\bigl (\nu S_2(\nu,Q^2)\bigr )-\bigl 
(\nu S^{pole}_2(\nu,Q^2)\bigr ) \bigr ] \nonumber \\
=&& \hspace {-0.5cm} 2\alpha I_2(Q^2) - \frac{2\alpha}{Q^2}
\bigl (I_{TT}(Q^2)-I_1(Q^2)\bigr )\nu^2 
+\frac{M^2}{Q^2} \gamma_{g_2}(Q^2) \nu^4 + O(\nu^6),
\end{eqnarray}
where the $O(\nu^4)$ term gives the generalized  $g_2$ polarizability:
\begin{eqnarray}
\gamma_{g_2}(Q^2)&=& \frac{16\pi \alpha M^2}{Q^6}\int_0^{x_0}x^2 g_2(x,Q^2)dx
\nonumber \\
=&&\delta_{LT}(Q^2)-\gamma_{TT}(Q^2)+\frac{2\alpha}{M^2 Q^2}
\Bigl (I_{TT}(Q^2)-I_1(Q^2)\Bigr ).
\end{eqnarray}

At high $Q^2$,
the OPE~\cite{SV,stein,Ji93,JU} for the VVCS amplitude leads to
the twist expansion:
\begin{equation}
\label{eq:Gamma1}
\Gamma_1(Q^2)
\equiv \int_0^1 g_1(x,Q^2)dx
=\sum_{\tau=2,4,...} \frac{\mu_\tau(Q^2)}{(Q^2)^{(\tau-2)/2}}
\end{equation}
with the coefficients $\mu_\tau$ related to nucleon matrix elements
of operators of twist $\leq \tau$.
Here twist is defined as the mass dimension minus the spin of an
operator, and the $\mu_\tau$ are perturbative 
series in $\alpha_s$, the running strong coupling constant.
Note that the application of the OPE requires a summation over all
hadronic final states including the ground state at $x=1$.

The leading-twist (twist-2) component, $\mu_2$, is determined by
matrix elements of the axial vector operator
$\bar\psi \gamma_\mu \gamma_5 \psi$ summed over quark flavors, where
$\psi$ are the quark field operators.
It can be decomposed into flavor triplet ($g_A$), octet ($a_8$) and
singlet ($\Delta\Sigma$) axial charges,
\begin{eqnarray}
\label{eq:mu2}
\mu_2(Q^2)
&=& 
  \left(\pm \frac{1}{12} g_A\ +\frac{1}{36} a_8 \right)
+ \frac{1}{9} \Delta\Sigma +O(\alpha_s(Q^2)) ,
\end{eqnarray}
where +(-) corresponds to proton (neutron), and
the $O(\alpha_s)$ terms involve the $Q^2-$evolution due to the QCD radiative
effects that can be calculated from perturbative QCD.
The triplet axial charge is obtained from neutron
$\beta$-decay, 
while the octet axial
charge can be extracted from hyperon weak-decay matrix elements assuming
SU(3) flavor symmetry. 
Within the quark-parton model $\Delta\Sigma$ is 
the amount of nucleon spin carried by the quarks. Deep Inelastic Scattering 
(DIS) experiments 
at large $Q^2$ have extracted this quantity through a global analysis 
of the world data~\cite{KCL09}. 

Eqs. (\ref{eq:Gamma1}) and (\ref{eq:mu2}), at leading twist and with the assumptions of SU(3) flavor symmetry and an
unpolarized strange sea, lead to the Ellis-Jaffe sum rule~\cite{EJ}.
The difference between the proton and the neutron gives the flavor non-singlet
term:
\begin{equation} 
\Gamma_1^p(Q^2)-\Gamma_1^n(Q^2)=\frac{1}{6}g_A+O(\alpha_s)+O(1/Q^2),
\label{eq:genBj}
\end{equation}
which becomes the Bjorken sum rule at the $Q^2\rightarrow \infty$ 
limit. 

If the nucleon mass were zero, $\mu_\tau$ would contain only a twist-$\tau$ operator. The non-zero nucleon mass induces contributions to $\mu_\tau$ from 
lower-twist operators.
The twist-4 term contains
a twist-2 contribution,
$a_2$, and a twist-3 contribution, $d_2$, in addition to $f_2$, the 
twist-4 component\cite{Ji93,JU,JM}:
\begin{equation}
\label{eq:mu4}
\mu_4
=M^2
\left( a_2 + 4 d_2 + 4 f_2 \right)/9 .
\end{equation}
The twist-2 matrix element $a_2$ is 
\begin{eqnarray}
 \label{eq:a2op}
 a_2\ S^{\{ \mu} P^\nu P^{\lambda\} }
 &=& \frac{1}{2} \sum_q e_q^2\
     \langle P,S |
    \bar\psi_q\ \gamma^{\{ \mu} iD^\nu iD^{\lambda\} } \psi_q
     | P,S \rangle\ ,
 \end{eqnarray}
where $e_q$ is the electric charge of a quark with flavor $q$, 
$D^\nu$ are the 
covariant derivatives and the parentheses
 $\{ \cdots \}$ denote symmetrization of indices.
The matrix element $a_2$ is related to the second moment of the twist-2 part 
of $g_1$:
\begin{equation}
 \label{eq:a2}
 a_2(Q^2)
 = 2 \int_0^1 dx\ x^2\ g_1(x,Q^2)\ .
 \end{equation}

Taking Eq.~(\ref{eq:a2}) as the definition of $a_2$, it is now generalized to any $Q^2$ including twist-2 and higher-twist contributions.
At low $Q^2$, the inelastic part of $a_2$ is related to 
$\gamma_{g_1}$, which is the generalized $g_1$ polarizability:
\begin{equation}
\label{eq:a2g}
{\overline {a_2}}(Q^2)=\frac{Q^6}{8\pi \alpha M^3}\gamma_{g_1}.
\end{equation}
Note that at large $Q^2$, the elastic contribution is negligible and
${\overline a_2}$ becomes $a_2$. 

The twist-3 component, $d_2$, is defined by the matrix element\cite{Ji93,JU,JM}:
\begin{equation}
 \label{eq:d2op}
 d_2 S^{[ \mu} P^{\{\nu ]} P^{\lambda\} }
 = \frac{1}{8} \sum_q e^2_q\
     \langle P,S |
    g \bar\psi_q\ \widetilde F^{\{ \mu \nu} \gamma^{\lambda\} } \psi_q
     | P,S \rangle\ ,
 \end{equation}
where $g$ is the QCD coupling constant,
 $\widetilde F^{\mu\nu}=(1/2)e^{\mu\nu\alpha\beta}F_{\alpha\beta}$ is the dual gluon-field strength tensor,
$F_{\alpha\beta}$ are the gluon field operators and the parentheses
 $[ \cdots ]$ denote antisymmetrization of indices.
This matrix element depends explicitly on gauge (gluon) fields. The gauge fields can be replaced by quark fields using the QCD equation of motion~\cite{JU}. 
Then, the matrix element can be related to the second moments of the twist-3 part of $g_1$ and $g_2$:
\begin{eqnarray}
 \label{eq:d2}
 d_2(Q^2)
&=& \int_0^1 dx\ x^2 \Bigl (2g_1(x,Q^2)+3g_2(x,Q^2)\Bigr ) \nonumber \\
=&&\hspace {-3mm} 3 \int_0^1 dx\ x^2\Bigl (g_2(x, Q^2)-g_2^{WW}(x,Q^2)\Bigr ),
 \end{eqnarray}
where $g_2^{WW}$ is the twist-2 part of $g_2$ as derived by Wandzura and 
Wilczek\cite{WW}
\begin{equation}
\label{eq:g2ww}
g_2^{WW}(x,Q^2)=-g_1(x,Q^2)+\int_x^1 dy \frac{g_1(y,Q^2)}{y}\ .
\end{equation}
The definition of $d_2$ with Eq.~(\ref{eq:d2}) is generalized to 
all $Q^2$.
At low $Q^2$,  the inelastic part of $d_2(Q^2)$ is related to the 
polarizabilities:
\begin{equation}
\label{eq:d2in}
{\overline {d_2}}(Q^2)
=\frac{8\pi \alpha M^3}{Q^6}(\gamma_{g_1}+\frac{3}{2}
\gamma_{g_2}) 
= 
\frac{Q^4}{8 M^4} \Bigl (I_1(Q^2)-I_{TT}(Q^2)+\frac{M^2Q^2}{\alpha}
\delta_{LT}(Q^2)\Bigr ) .
\end{equation}
At large $Q^2$, ${\overline {d_2}}$ becomes $d_2$ since the elastic
contribution becomes negligible.

The twist-4 contribution to $\mu_4$ is defined by the
matrix element
\begin{eqnarray}
\label{eq:f2op}
f_2\ M^2 S^\mu
&=& \frac{1}{2} \sum_q e_q^2\
\langle P,S |
 g \bar\psi_q\ \widetilde{F}^{\mu\nu} \gamma_\nu\ \psi_q
| P,S \rangle\ .
\end{eqnarray}

This matrix element depends also explicitly on gauge (gluon) fields. The 
gluon fields can be replaced with the ``bad'' components of quark fields
using the QCD equation of motion~\cite{Ji93,JU}. Then this matrix element can be related 
to moments of parton distributions:
\begin{equation}
 \label{eq:f2}
 f_2(Q^2) = \frac{1}{2}\int_0^1 dx\ x^2 
\Bigl (7g_1(x,Q^2)+12g_2(x,Q^2)-9g_3(x,Q^2)\Bigr ), 
 \end{equation}
where 
$$g_3(x,Q^2)=-\frac{1}{M^2}\int_0^1 \frac{d\lambda}{2\pi }e^{i\lambda x}
\sum_q e^2_q \langle P,S | \bar\psi_q\ \not p \gamma_5 \
\psi_q\ 
| P,S \rangle\
$$
is the twist-4 distribution function defined in terms of the 
``bad'' components of quark fields in the light-cone gauge, 
and $p$ is along the
direction of ``bad'' components (perpendicular to the light cone 
direction)~ 
\footnote {This $g_3$ is not to be confused with the parity-violating 
structure function, which also uses the notion of $g_3$ in the literature.}.
With only $g_1$ and $g_2$ data available, $f_2$ can be extracted through 
Eqs. (\ref{eq:Gamma1}) and (\ref{eq:mu4}) if the twist-6 or higher terms are
not significant or are included in the extraction. 

The twist-3 and 4 operators describe the response of the collective
color electric and magnetic fields to the spin of the nucleon.
Expressing these matrix elements in terms of the components of
$\widetilde{F}^{\mu\nu}$ in the nucleon rest frame, one can relate
$d_2$ and $f_2$ to color electric and magnetic polarizabilities.
These are defined as~\cite{JU,JM}
\begin{equation}
\chi_E\ 2 M^2 \vec S
= \langle N |\ \vec j_a \times \vec E_a\ | N \rangle\ , \ \ \
%
\chi_B\ 2 M^2 \vec S
= \langle N |\ j_a^0\ \vec B_a\ | N \rangle\ ,
\end{equation}
where $\vec S$ is the nucleon spin vector,
$j_a^\mu$ is the quark current,
$\vec E_a$ and $\vec B_a$ are the color electric and
magnetic fields, respectively.
In terms of $d_2$ and $f_2$ the color polarizabilities can be
expressed as
\begin{equation}
\chi_E = \frac{2}{3} \left( 2 d_2\ +\ f_2 \right),  \ \ \
\chi_B = \frac{1}{3} \left( 4 d_2\ -\ f_2 \right).
\label{eq:chi}
\end{equation}

Recently, M. Burkardt pointed out~\cite{burkardt} that $d_2$ and $f_2$ are 
local correlators. He suggested that instead of calling them color 
polarizabilities, they may more appropriately be identified with the 
transverse component of the color-Lorentz force acting on the struck quark.
Furthermore, he pointed out an interesting link between $d_2$ and the 
Sivers function, a transverse-momentum-dependent (TMD) distribution function. 
Since this is beyond the scope of this article, interested readers can find 
the details in Ref.~\cite{burkardt}.

\section{Summary of experimental situation before JLab}
Before Jefferson Lab started experiments with polarized beams and targets,
 most of the spin structure 
measurements of the nucleon were performed at high-energy facilities like CERN (EMC~\cite{EMC} and SMC~\cite{SMCp,SMCd,SMCd2,SMCd0,SMCp2}), 
DESY (HERMES~\cite{HERMESn,HERMESGDH,HERMESp}) and SLAC (E80~\cite{E80}, E130~\cite{E130}, E142~\cite{E142},
E143~\cite{E143}, E154~\cite{E154,E154g2}, E155~\cite{E155,E155g2} and E155x~\cite{E155x}). 
The measured $g_1$ 
and $g_2$ data were suitable for an analysis in terms of perturbative QCD. The impetus for 
performing these experiments on both the proton and the neutron was to test the Bjorken 
sum rule, a fundamental sum rule of QCD. After twenty-five years of active 
investigation 
this goal was accomplished with a test of this sum rule to better than 10\%. 
The spin structure of the nucleon was unraveled in the same process. 
Among the 
highlights of this effort is the determination of the total spin content of the nucleon due to 
quarks, $\Delta\Sigma$ (see Eq.~(\ref{eq:mu2})). This study also revealed the 
important role
of the quark orbital angular momentum and gluon total angular momentum. 
The main results
from this inclusive double spin asymmetry measurement program have led to new directions: namely the 
quest for an experimental determination of the orbital angular momentum contribution~\cite{ji97,Bur05} (e.g., with Deep Virtual Compton Scattering 
and Transverse Target Single Spin Asymmetry measurements 
at Jefferson Lab, HERA and CERN) and
the gluon spin contribution (with  COMPASS~\cite{COMPASS} and  RHIC-spin~\cite{RHICspin} experiments). These efforts will be ongoing 
for the next few decades.

\section{Recent results from Jefferson Lab}

\subsection{JLab experiments}
\label{JLabexp}

With a high current, high polarization electron beam of energy up to 6 GeV
and state-of-the-art polarized targets, JLab has completed a number of 
experiments,
which have extended the database on spin structure functions significantly 
both in kinematic range (low $Q^2$ and high $x$) and in precision. 
The neutron results are from Hall A\cite{HallA nim} using polarized 
$^3$He\cite{Astatus07} as an effective 
polarized neutron target and two 
high resolution spectrometers. The polarized luminosity reached 10$^{36}$
s$^{-1}$cm$^{-2}$ with in-beam polarization improved from $35\%$ (1998) to over 
$65\%$ (2008).
The proton and deuteron results are from Hall B\cite{HallB nim} with the
CLAS detector and Hall C with the HMS spectrometer and using polarized 
NH$_3$ and ND$_3$ targets\cite{NH3} 
with in-beam polarization of about $80\%$ and $40\%$, respectively.
\subsection{Results of the generalized GDH sum for the neutron and $^3$He}
\label{GDHsum}

\begin{figure}[ht!]
\begin{center}
\includegraphics[scale=0.45,angle=0]{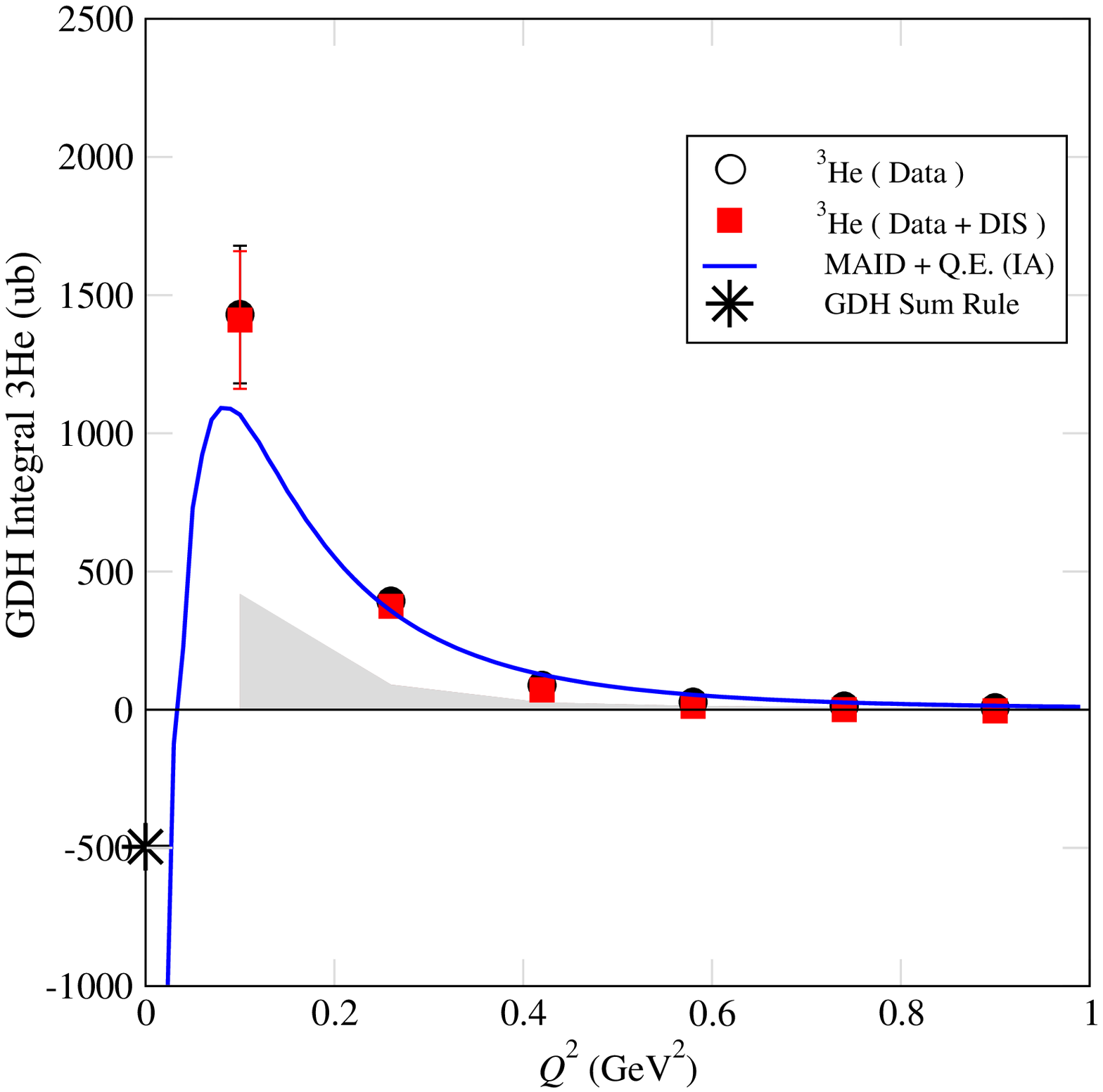}
\includegraphics[scale=0.45,angle=-90]{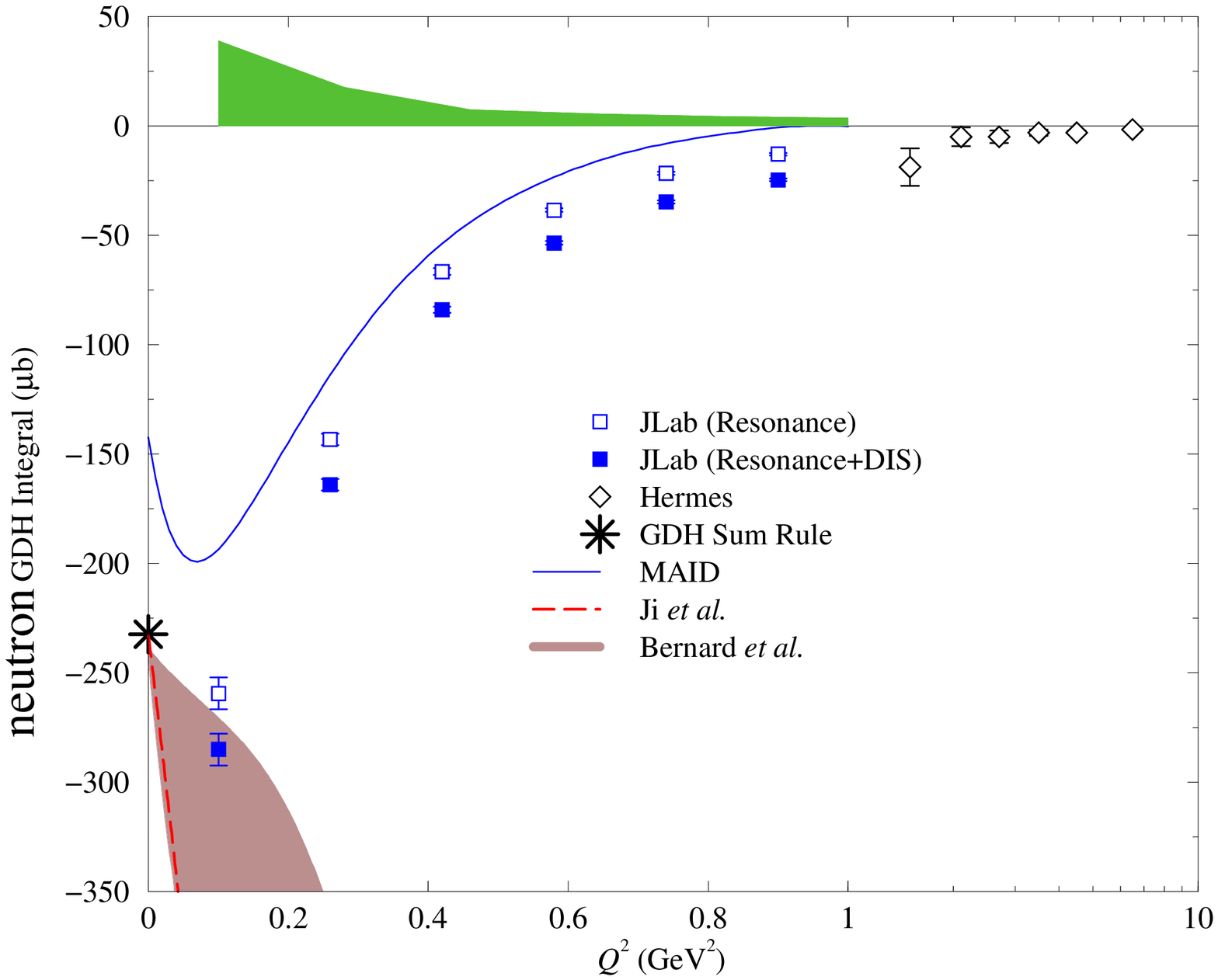}
\caption{Results for the GDH sum $I(Q^2)$ for $^3$He (top) and
 the neutron (bottom). The $^3$He GDH results are 
compared with the MAID model plus a quasielastic contribution. The neutron GDH results are compared with 
$\chi$PT calculations of ref.~\protect\cite{ji00} (dotted curve) and ref.~\protect\cite{ber03} 
(dot-dashed curve). The MAID model calculation of ref.~\protect\cite{dre01},
is represented by a solid curve.  Data from HERMES~\protect\cite{HERMESGDH} are also shown.}
\label{fig:GDH}
\end{center}
\end{figure}

The spin structure functions $g_1$ and $g_2$ (or $\sigma_{TT}$ and 
$\sigma_{LT}$) were measured in Hall A experiment E94-010 on $^3$He from
break-up threshold to $W=2$ GeV covering the $Q^2-$range of 0.1-0.9 GeV$^2$.
The generalized GDH integrals  
$I(Q^2)$ (Eq.\ref{eq:gdhsum_def1})
(open symbols) were extracted for $^3$He~\cite{e94010-He3} (top plot of 
Fig.~\ref{fig:GDH}) and for the neutron~\cite{e94010-1} 
(bottom). The solid squares include an estimate of the 
unmeasured high-energy part.
  
 The $^3$He results rise with decreasing $Q^2$. Since the GDH sum 
rule at $Q^2=0$ predicts a large negative value, a drastic turn around 
should happen at $Q^2$ lower than 0.1 GeV$^2$. A simple model using MAID 
plus quasielastic contributions estimated from a PWIA model~\cite{cio97} indeed shows the expected turn
around. The data at low $Q^2$ should be a good testing ground for few-body chiral perturbation theory calculations when they are available.

The neutron results indicate a significant yet smooth variation of $I(Q^2)$  
to increasingly negative values as $Q^2$ varies from $0.9\,{\rm GeV^2}$ 
towards zero.
The data are more negative than the MAID model calculation~\cite{dre01}. 
Since the calculation only includes contributions to $I(Q^2)$ for $W \le 2\,{\rm GeV}$, 
the model should be compared with the open squares.
The GDH sum rule 
prediction, $I(0)=-232.8\,\mu{\rm b}$, is indicated along with  
extensions to $Q^2>0$ using two next-to-leading order chiral perturbation 
theory ($\chi$PT)
calculations: one of them using the heavy baryon approximation (HB$\chi$PT)~\cite{ji00} 
(dotted line) and the other relativistic baryon $\chi$PT (RB$\chi$PT)~\cite{ber03} (dot-dashed line). Shown with a brown band is a RB$\chi$PT prediction 
including resonance effects~\cite{ber03}. The
large uncertainty is due to the resonance parameters used. 

Improved calculations\cite{krebs09,kao08} as well as further measurements~\cite{e97110}
will help clarify the situation. 
\subsection{First moments of $g_1$ and the Bjorken sum}
\begin{figure}[ht!]
\begin{center}
\includegraphics[scale=0.34]{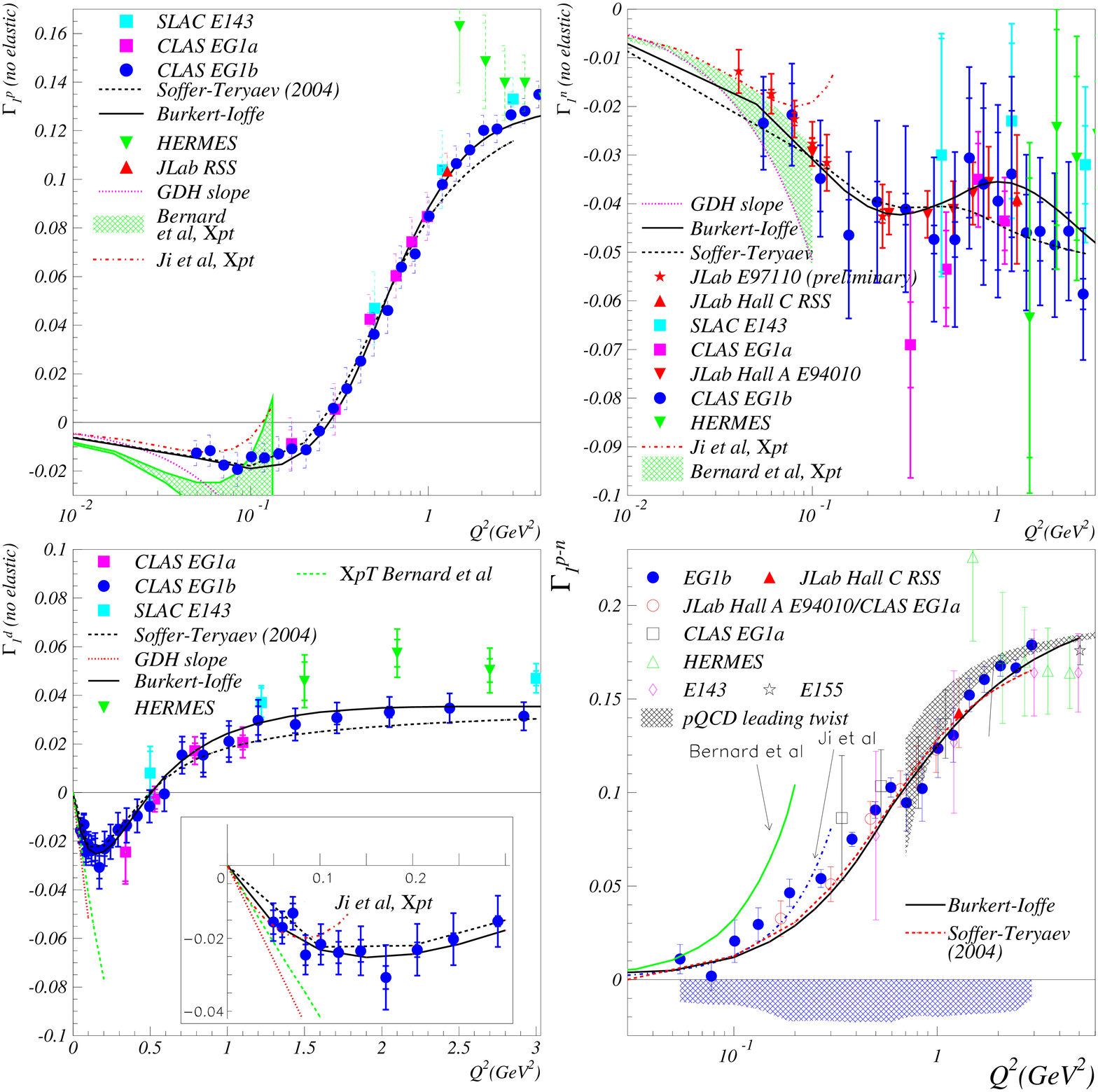}
\caption {Results of $\Gamma_1 (Q^2)$ for p, d, n and p-n from the JLab Hall A,
CLAS eg1 and Hall C RSS experiments.
The slopes at $Q^2=0$ predicted by the GDH sum rule are given by the dotted 
lines. The dashed (plain) lines are the predictions from the 
Soffer-Teryaev~\protect\cite{sof02} (Burkert-Ioffe~\protect\cite{bur92} ) model.
The leading twist $Q^2$-evolution of the moments is given by the gray band. 
The inset shows comparisons with $\chi$PT calculations by 
Ji \emph{et al.}~\protect\cite{ji00} and Bernard \emph{et al.}~\protect\cite{ber03}. } 
\label{fig:gamma1pn}
\end{center}
\end{figure}

The first moment of $g_1$, $\Gamma_1$, was extracted from E94-010 
for the neutron~\cite{e94010-2} and $^3$He~\cite{e94010-He3}, and  
from the CLAS eg1 experiment for the 
proton~\cite{eg1p}, the deuteron~\cite{eg1d} and the neutron 
(from the deuteron with the proton contribution subtracted)
over a $Q^2-$range from 0.05 to 5 GeV$^2$. The results are plotted in 
Fig.~\ref{fig:gamma1pn}. Also plotted are the results at $Q^2 = 1.3$ GeV$^2$ 
from the Hall C RSS experiment~\cite{RSS}.
These moments show a strong yet smooth 
variation in the transition region ($Q^2$ from 1 to 0.05 GeV$^2$).
The lowest $Q^2$ (0.05-0.1 GeV$^2$) data are compared with two $\chi$PT calculations and are in reasonable agreement.

Also plotted are the preliminary neutron data at very low $Q^2$ of 
0.04 to 0.24 GeV$^2$ from the Hall A E97-110 experiment. The precision of the
preliminary data is  currently limited by systematic uncertainties, which 
are expected to be reduced significantly when the final results become 
available. The new data are in good agreement with published results in the 
overlap region. The data in the very low $Q^2-$region provide a benchmark 
test of the $\chi$PT calculations, since they are expected to work in this 
region. The results agree well with both $\chi$PT calculations and also 
indicate a smooth transition as $Q^2$ approaches zero.
 
The bottom-right panel shows the 
first moment for p-n~\cite{BJ}, a flavor non-singlet combination, which 
is the Bjorken sum at large $Q^2$~[\ref{eq:genBj}].
The Bjorken sum was used to extract the strong coupling constant $\alpha_s$
at high $Q^2$ (5 GeV$^2$). An attempt was made
to extract an ``effective coupling'' $\alpha_{s,g_1}$ in the 
low $Q^2-$region using
the Bjorken sum, see section~\ref{alpha_s}. 

At large $Q^{2}$, the data agree with pQCD models
where the higher-twist effects are small.

\subsection{First moment of $g_2$: B-C sum}

Measurements of $g_2$ require transversely polarized targets. SLAC E155x~\cite{E155x} performed the only dedicated $g_2$ measurement prior to JLab. At JLab, $g_2$ and its moments have been extensively measured for the neutron with a polarized $^3$He target in a wide range of kinematics in several Hall A experiments (E94-010~\cite{e94010-2}, E97-103~\cite{e97103}, E99-117~\cite{Zheng}, E02-012~\cite{e02012-2} and E97-110~\cite{e97110}). The measurement on the proton was performed in the RSS experiment in Hall C at an average $Q^2$ of 1.3~GeV$^2$.

\begin{figure}
\begin{center}
\includegraphics[scale=0.6]{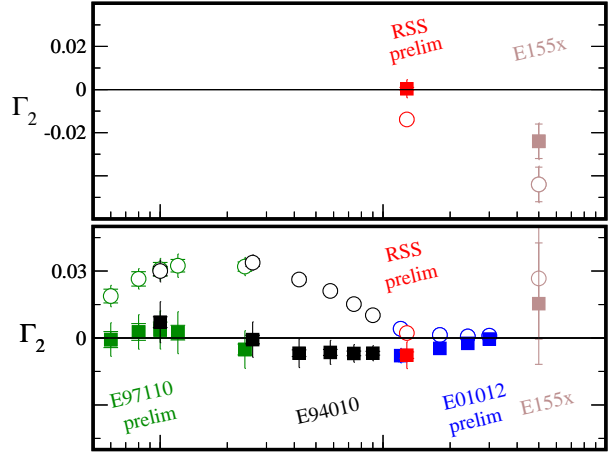}
\end{center}
\caption{\label{gamma2} The verification of the BC sum rule from Hall C RSS and Hall A experiments
E94-010, E97-110 and E02-012 (neutron only), together with SLAC E155x data. The top plot is for the proton and the bottom for the neutron. The open symbols are the measured values and the solid symbols are the total moments, including 
the elastic and estimated contributions from the unmeasured high-energy region.}
\end{figure}

The first moment of $g_2$ is expected to be zero at all $Q^2$ from the 
Burkhardt-Cottingham (B-C) sum rule. The SLAC E155x results yielded a first test of 
the B-C sum rule for the proton, deuteron and neutron (extracted from p and d 
data).
The proton result (top panel of Fig.~\ref{gamma2}) appears to be 
inconsistent with the B-C sum rule at the 
2.75 $\sigma$ level. In addition to the large experimental uncertainty, there 
is an uncertainty associated with the low-$x$ extrapolation that is difficult to
quantify. The deuteron and neutron (bottom panel of Fig.~\ref{gamma2}) results 
from SLAC show agreement with the 
B-C sum rule with large uncertainties.
The most extensive measurements of the B-C sum rule come from 
experiments~\cite{e97110,e94010-2,e02012-2} on 
the neutron using a longitudinally and transversely polarized $^3$He target 
in Hall A. The results for $\Gamma_2^n$ are plotted in the bottom panel of 
Fig.~\ref{gamma2} in the measured region (open circles). The solid diamonds 
include the elastic contribution and an estimated DIS contribution assuming
$g_2=g_2^{WW}$. The published results from E94-010 and the 
preliminary results from
E97-110 and E02-012 are the most precise data on the B-C sum rule and are 
consistent with the expectation of zero, within small systematic uncertainties. 
The RSS experiment in Hall C~\cite{RSS} took data on $g_2$ for the proton and 
deuteron, at an average $Q^2$ of about 1.3 GeV$^2$. For the proton (top panel
of Fig.~\ref{gamma2}), the 
integral over the measured resonance region is negative, but after the elastic 
contribution and the estimated small-$x$ part are added, the preliminary 
result for the 
B-C sum is consistent with zero. Preliminary result on the deuteron and the 
neutron (from d and p) are also 
consistent with zero within uncertainties.
The precision data from JLab are consistent with the B-C sum rule in
all cases, indicating that $g_2$ is a well-behaved function 
with good convergence at high energy 
though the high energy part is mostly unmeasured.
 
\subsection{Spin Polarizabilities: $\gamma_0$, $\delta_{LT}$}
The generalized spin polarizabilities provide benchmark tests
of $\chi$PT calculations at low $Q^2$.
Since the generalized polarizabilities have an extra $1/\nu^2$ 
weighting compared to the first moments (GDH sum or 
$I_{LT}$), these integrals have less of a contribution from the large-$\nu$ 
region and converge much faster, which minimizes the uncertainty due to
the unmeasured region at large $\nu$. 

At low $Q^2$, the 
generalized polarizabilities have been evaluated with $\chi$PT 
calculations~\cite{ber02,van02}.
One issue in the $\chi$PT calculations is how to properly
include the nucleon resonance contributions, especially the $\Delta$ resonance
which usually dominates.
As was pointed out in Ref.~\cite{ber02,van02}, while $\gamma_0$ is sensitive to 
resonances, $\delta_{LT}$ is insensitive to the $\Delta$ 
resonance. Measurements of the generalized spin
polarizabilities are an important step in understanding the dynamics of
QCD in the chiral perturbation region.

The first results for the neutron generalized forward
spin polarizabilities $\gamma_0(Q^2)$ and $\delta_{LT}(Q^2)$
were obtained at Jefferson Lab Hall A~\cite{e94010-3}
over a $Q^2-$range from 0.1 to 0.9 GeV$^2$.
We will focus on the low $Q^2-$region where the comparison with $\chi$PT
calculations is meaningful.

\begin{figure}[ht!]
\begin{center}
\centerline{\includegraphics[scale=0.22,angle=0]{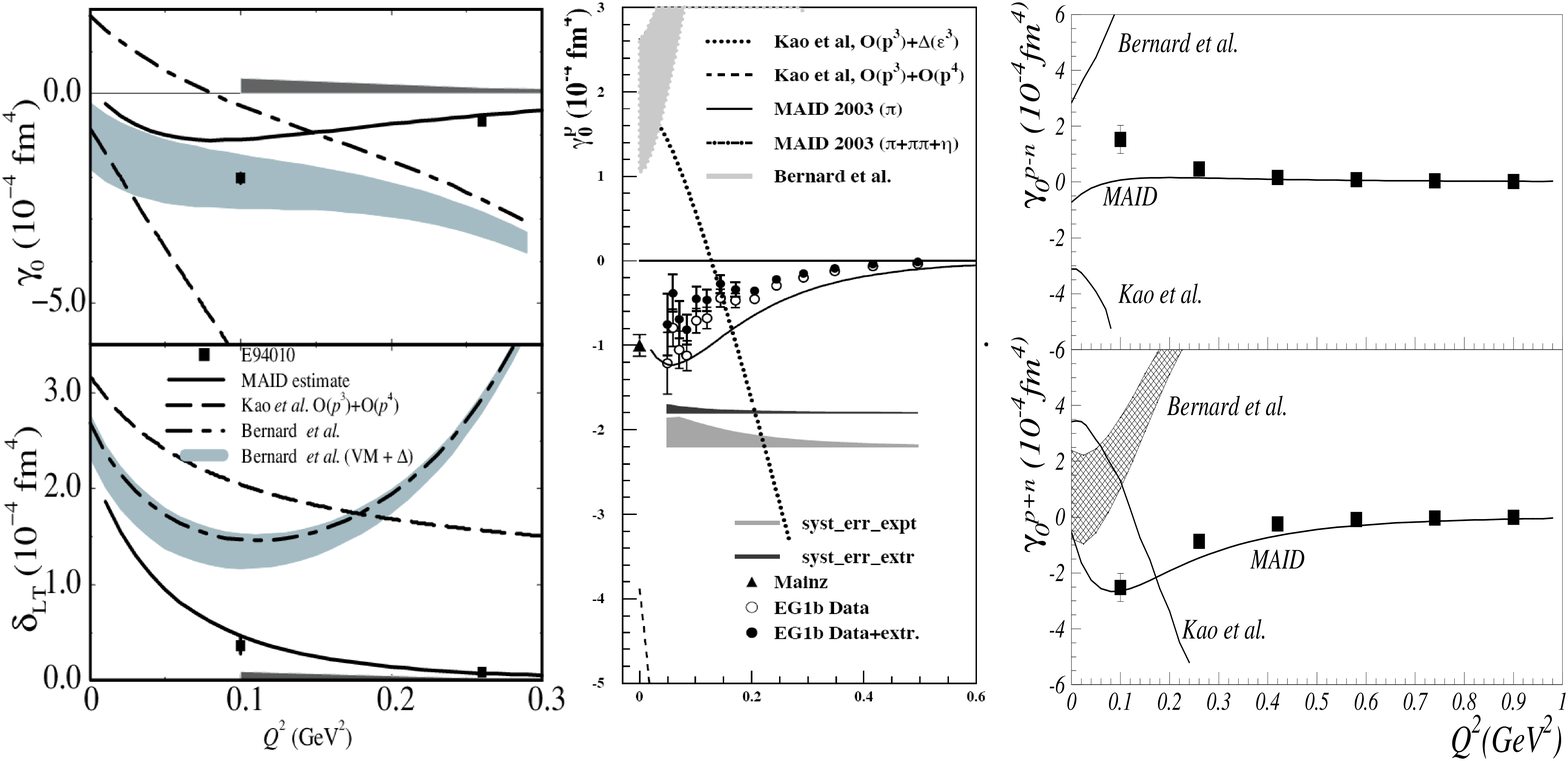}}
\end{center}
\caption{Left panels display results for the neutron spin polarizabilities 
$\gamma_0$ (top) and $\delta_{LT}$
(bottom). Solid squares are the results with statistical uncertainties.
The dark bands indicates the systematic uncertainties. 
The dashed curves are the HB$\chi $PT 
calculation\protect\cite{van02}. 
The dot-dashed curves and the light bands are the 
RB$\chi $PT calculation\protect\cite{ber02} 
without and with\protect\cite{ber03} 
the $\Delta $ and vector-meson contributions, respectively.
Solid curves are from the MAID model\protect\cite{dre01}.
The other panels are results for the spin polarizability 
$\gamma_0$ for the proton, $p-n$ and $p+n$.
}
\label{fig:gam0dlt}
\end{figure}

The results for $\gamma_0(Q^2)$ for the neutron are shown
in the top-left panel of Fig.~\ref{fig:gam0dlt} for the two 
lowest $Q^2-$values of 0.10 and 0.26 GeV$^2$. 
The statistical uncertainties are smaller than the
size of the symbols. 
The data are compared with 
a next-to-leading order, $O(p^4)$, HB$\chi$PT 
calculation\cite{van02}, a next-to-leading-order RB$\chi$PT
calculation\cite{ber02} and the same calculation explicitly including 
both the $\Delta$ resonance and vector-meson contributions\cite{ber03}.
Predictions from the MAID model\cite{dre01} are also shown.
At the lowest $Q^2$ point,
the RB$\chi$PT
calculation including the resonance contributions
agrees with the experimental result.
For the HB$\chi$PT calculation without explicit resonance contributions, 
the discrepancy is large even at $Q^2 = 0.1$ GeV$^2$. 
This might indicate the significance of the resonance contributions or a
problem with the heavy baryon approximation at this $Q^2$.
The higher-$Q^2$ data point is in good agreement with the MAID
prediction,
but the lowest data point at $Q^2 = 0.1 $ GeV$^2$ is significantly lower, 
consistent with what was observed for the generalized GDH
integral result (Section~\ref{GDHsum}).
Since $\delta_{LT}$ is 
insensitive to the dominating
$\Delta$-resonance contribution, it was believed that $\delta_{LT}$ should be
more suitable than $\gamma_0$ to serve as a testing ground for the chiral 
dynamics of QCD\cite{ber02,van02}.
The bottom-left panel of Fig.~\ref{fig:gam0dlt} shows $\delta_{LT}$ 
compared to
$\chi$PT calculations and the MAID predictions. It is surprising to see
that 
the data are in significant disagreement with the $\chi$PT calculations 
even at the lowest-$Q^2$ of 0.1 GeV$^2$. 
This discrepancy presents a significant challenge to the present theoretical
understanding. 
The MAID predictions are in good agreement with the results.

The other panels of 
Fig.~\ref{fig:gam0dlt} present results of
$\gamma_{0}$ for the proton~\cite{eg1gam0}
and for the isospin decompositions: p-n and p+n~\cite{BJ}.

These results are in strong disagreements 
with both $\chi$PT calculations. Since $\gamma_{0}$ is sensitive to the 
$\Delta$ contributions, this may indicate that the treatment of
the $\Delta$ contributions in the $\chi$PT calculations needs 
to be taken into account properly. Progress in this direction is expected
in the near future~\cite{krebs09}.

\subsection{Higher moment $d_2$ and higher twist: twist-3}
\label{d2}

\begin{figure}[ht!]
\begin{center}
\vspace{5mm}
\epsfig{file=d2n.eps, scale=0.45}
\epsfig{file=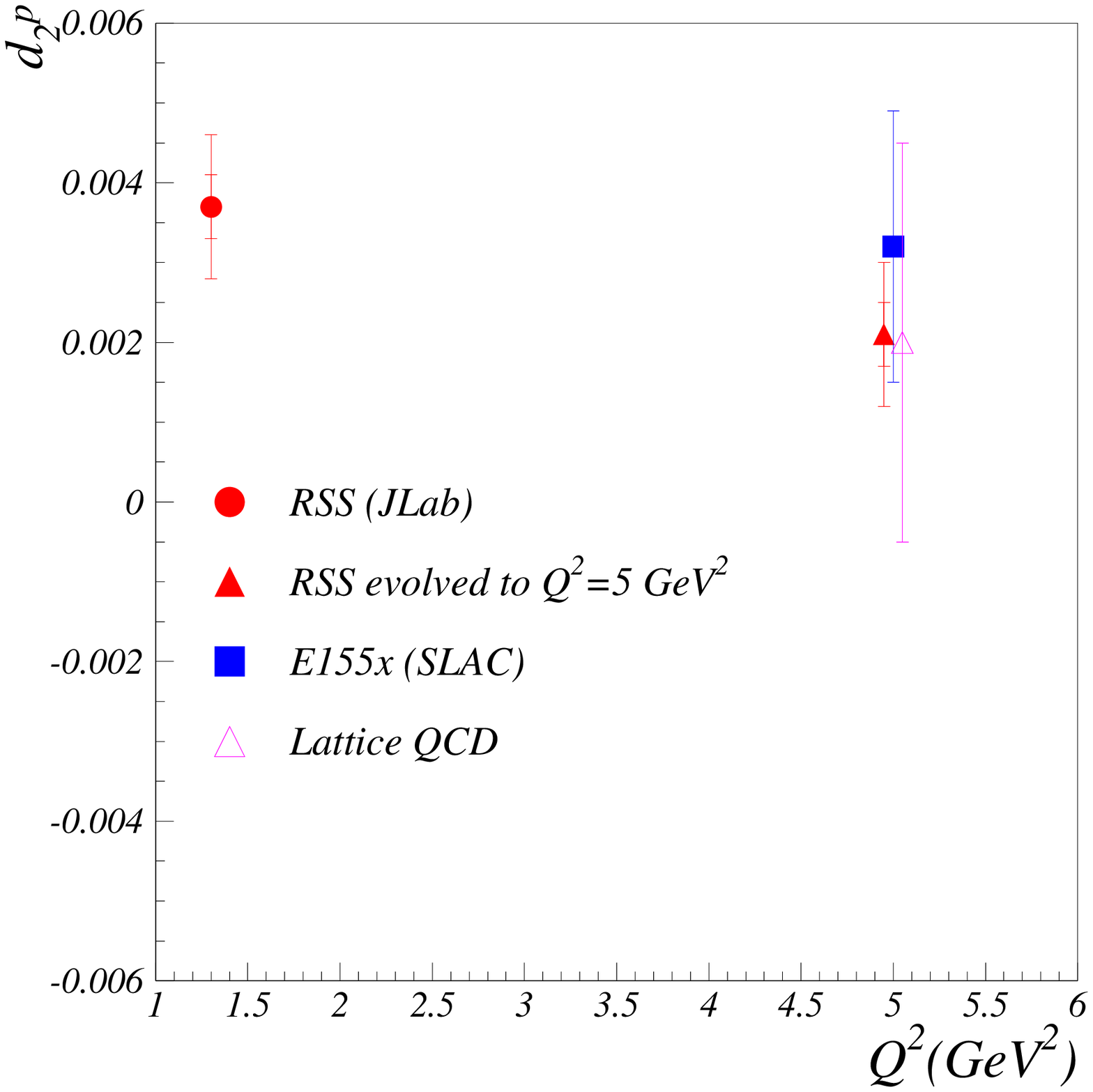, scale=0.30}
\end{center}
\caption{The left plot shows the Hall A results of $\overline d_2$ for the 
neutron 
along with the world data at high $Q^2$, Lattice QCD, MAID model and HB$\chi$PT
calculations. The right plot shows the Hall C result of $\overline d_2$ for 
the proton at $Q^2=1.3$ GeV$^2$, the same data evolved to $Q^2=5 $ GeV$^2$
along with the SLAC data and lattice QCD calculation.
}
\label{fig:d2}
\vspace{-5mm}
\end{figure}

Another combination of the second moments, $d_2(Q^2)$, provides an 
efficient way to study the high-$Q^2$ behavior of nucleon spin structure,
since at high $Q^2$ it is related to a matrix element $d_2$ (color 
polarizability) which can 
be calculated from Lattice QCD. This moment also provides a means to study the 
transition from high to low $Q^2$. 
In the left panel of Fig.~\ref{fig:d2}, $\bar d_2(Q^2)$  
on the neutron is shown. The preliminary results from E02-012~\cite{e02012-2} 
and the 
published E94-010~\cite{e94010-2} results are shown 
as the solid squares and open circles, respectively. 
The bands represent the systematic uncertainties. The neutron results 
from SLAC\cite{E155x}(open diamond) at high $Q^2$ 
and from combined SLAC and JLab E99-117\cite{Zheng}(solid diamond) 
are also shown. The solid line is the
MAID calculation containing only the resonance contribution. 
At moderate $Q^2$,
our data show that $\bar d_2^n$ is positive and decreases with $Q^2$.
At large $Q^2$ ($Q^2 = 5$~GeV$^2$), data (SLAC alone and SLAC 
combined with JLab E99-117) are positive.  
The lattice QCD prediction\cite{goc01} at $Q^2 = 5$~GeV$^2$ is 
negative but close to zero and is about two sigmas away from 
the data. 
We note that all models (not shown at
this scale) predict a small negative or zero value at large $Q^2$. 
High-precision data at large $Q^2$ will be crucial for a benchmark test of the
lattice QCD predictions and for understanding the dynamics of the quark-gluon 
correlations. 

Also shown are the $d_2$ result on the proton from Hall C~\cite{RSS-2} at 
$Q^2=1.3$ GeV$^2$ and the same result evolved to $Q^2=5 $ GeV$^2$ along 
with the SLAC data and the Lattice
QCD calculation. 

\subsection{Higher twist extractions: $f_2$, twist-4}
\label{HT}
The higher-twist contributions to $\Gamma_1$ can be obtained
by a fit with an OPE series, Eq.~(\ref{eq:Gamma1}), truncated to an order
appropriate for the precision and $Q^2$-span of the data.
The goal is to determine the twist-4 matrix element $f_2$.
Once $\mu_4$ is obtained, $f_2$ is extracted 
by subtracting the leading-twist contributions of $a_2$ and $d_2$ 
following Eq.~(\ref{eq:mu4}). The data precision and $Q^2-$range allow
a fit to both the $\mu_4$ and $\mu_6$ terms.
To have an idea how the higher-twist terms
(twist-8 and above) affect the twist-4 term extraction,
it is necessary to study the convergence of the expansion 
and to choose the $Q^2-$range in a region where the $\mu_8$ term is not
significant. A fit with three terms ($\mu_4$, $\mu_6$ and $\mu_8$) 
was also preformed for this purpose.
This study was possible only because of the 
availability of the high-precision low-$Q^2$ data from JLab.

Higher-twist analyses 
have been performed on the proton\cite{HTP2}, the neutron\cite{HTN} and the Bjorken 
sum\cite{BJ}. An earlier proton analysis is available\cite{HTP} but will not 
be presented here, since that analysis used a different procedure.
$\Gamma_1$ at moderate $Q^2$ is obtained as described in section~\ref{JLabexp}.
For consistency, the unmeasured low-$x$ parts of the JLab $\Gamma_1^p$ and 
of the world data on $\Gamma_1$ were re-evaluated using the same prescription 
previously used for $\Gamma_1^n$ and $\Gamma_1^{p-n}$. 
The elastic contribution, negligible above 
$Q^2$ of 2 GeV$^2$ but significant (especially for the proton) at lower values 
of $Q^2$, was added using the parametrization of Ref.~\cite{mer96}.
The leading-twist term $\mu_2$ was determined by fitting the 
data at $Q^2 \ge 5$ GeV$^2$ assuming that higher twists are negligible
in this $Q^2-$region. A value of
$g_A=1.270 \pm 0.045$ was obtained for the Bjorken sum.
Using the proton (neutron) data alone, and with input of 
$g_A$ from neutron beta decay and $a_8$ from hyperon 
decay (assuming SU(3) flavor symmetry), 
we obtained $\Delta\Sigma=0.15 \pm 0.07$ for the proton and
$\Delta\Sigma=0.35\pm 0.08$ for the neutron.  We note that there 
is a significant difference between $\Delta\Sigma$ determined from the proton 
and from the neutron data.
This is the main reason why the extracted $\mu_4$
and $f_2$ from the Bjorken sum is different compared to the difference of those 
extracted individually from the proton and neutron, since the Bjorken sum does not 
need the assumption of SU(3) flavor symmetry and $\Delta \Sigma$ was canceled.

The fit results using an expansion up to ($1/Q^6$) in determining $\mu_4$ are 
summarized in Table 1. 
In order to extract $f_2$, as shown in Table 2,
the target-mass corrections $a_2$ were evaluated using the Blumlein-Boettcher world data parametrization\cite{BB} for the proton and a fit to the world 
neutron data, which includes the recent
high-precision neutron results at large $x$\cite{Zheng}. The $d_2$ values 
used are from SLAC E155x\cite{E155x} (proton) and JLab E99-117\cite{Zheng}
(neutron). 
\begin{table}[ht]
\tbl{Results of $\mu_4$, $\mu_6$ and 
$\mu_8$ at $Q^2$ = 1 GeV$^2$ for proton, neutron and $p-n$. The uncertainties are
first statistical, then systematic.} 
{\begin{tabular}{|c|c|c|c|c|}
\hline 
Target& $Q^{2}$ (GeV$^{2}$)&
$\mu_4/M^2$&
$\mu_6/M^4$&
$\mu_8/M^6$\tabularnewline
\hline
\hline 
proton & 0.6-11.0&
-0.065$\pm 0.012 \pm 0.048$&
0.143$\pm 0.021 \pm 0.056$&
-0.026$\pm 0.008 \pm 0.016$\tabularnewline
neutron&0.5-11.0&
0.019$\pm 0.002 \pm 0.024$&
-0.019$\pm 0.002 \pm 0.017$&
0.00$\pm 0.00 \pm 0.03$\tabularnewline
$p-n$&0.5-11.0&
-0.060$\pm 0.045 \pm 0.018$&
0.086$\pm 0.077 \pm 0.032$&
0.011$\pm 0.031 \pm 0.019$\tabularnewline
\hline
\end{tabular}}
\label{table:mu468}
\end{table}
\begin{table}[ht]
\tbl{Results of $f_2$, $\chi_E$ and 
$\chi_B$ at $Q^2$ = 1 GeV$^2$ for proton, neutron and $p-n$.
The uncertainties are
first statistical, then systematic.} 
{\begin{tabular}{|c|c|c|c|}
\hline 
Target & $f_2$ & $\chi_E$ & $\chi_B$ \tabularnewline
\hline
\hline 
$p$ & -0.160 $\pm 0.028 \pm 0.109$ & -0.082 $\pm 0.016 \pm 0.071$ &
0.056 $\pm 0.008 \pm 0.036$\tabularnewline
$n$ & 0.034 $\pm 0.005 \pm 0.043$ & 0.031 $\pm 0.005 \pm$ 0.028 & 
-0.003 $\pm$ 0.004 $\pm$ 0.014 \tabularnewline
$p-n$ & $-0.136 \pm 0.102 \pm 0.039$ & $-0.100 \pm 0.068 \pm 0.028 $
& $0.036 \pm 0.034 \pm 0.017 $ \tabularnewline
\hline
\end{tabular}}
\label{table:f2}
\end{table}

The $\Gamma_1$ moments were fit, varying the minimum $Q^2-$value
to study the convergence of the OPE series. 
The extracted quantities have large uncertainties (dominated by the 
systematic uncertainty) but are stable with respect to a minimal $Q^2-$value 
when it is below 1 GeV$^2$. 
The results do not vary significantly when the $
\mu_8$ term is added, which justifies \emph{a posteriori} the use of the 
truncated OPE series in the chosen $Q^2-$range. In the proton case, the 
elastic contribution makes a significant contribution to the $\mu_6$ term at 
low $Q^2$ but this does not invalidate \emph{a priori} the validity of the
series since the elastic contributes mainly to $\mu_6$ and $\mu_8$, but remains
small compared to $\mu_4$. We notice the alternation of signs between the 
coefficients. This leads to a partial suppression of the higher-twist 
effects and may be a reason for quark-hadron duality in
the spin sector\cite{duality}. We also note that the sign 
alternation is opposite for the proton and neutron. 
Following Eq.~(\ref{eq:chi}), the electric and magnetic color polarizabilities were determined. 
Overall, the values given 
in Table 2 are small, and we observe a sign change 
in the electric color polarizability between the proton and the neutron. 
We also expect a sign change in the color magnetic polarizability. 
However, with the large uncertainty and the 
small negative value of the neutron $\chi_B$, it is difficult to confirm 
this expectation.

\subsection{Quark-hadron duality in spin structure functions}
\label{duality}

\begin{figure}[htb!]
\begin{center}
\epsfig{file=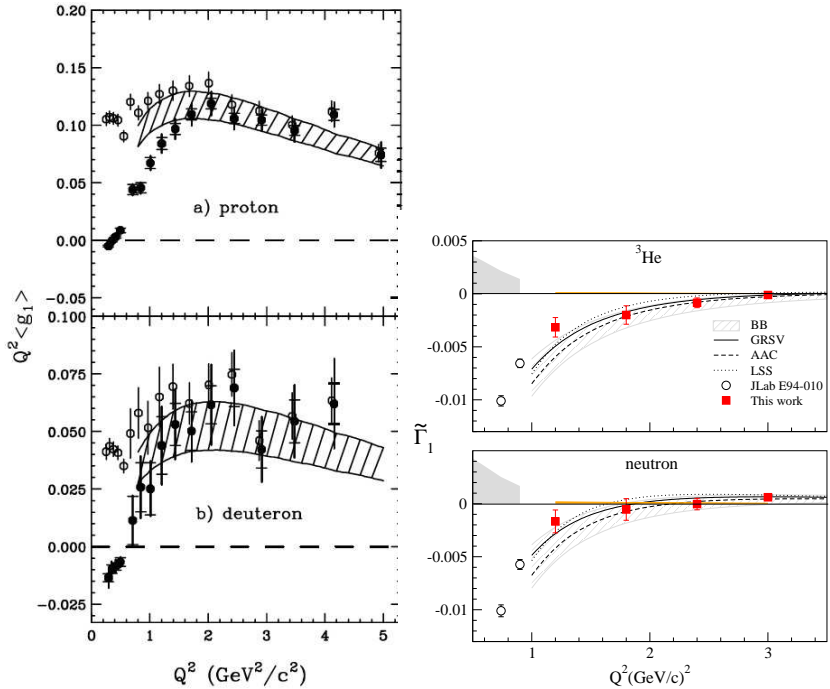, scale=0.34}
\epsfig{file=gamma1_duality.eps, scale=0.38}
\caption{Data on $g_1(x,Q^2)$ averaged over the resonance region for the 
proton and the deuteron from Hall B eg1 (left panel).
The hatched curves
represent the range of extrapolated DIS results from modern NLO fits (GRSV and AAC), evolved to the $Q^2$ of the data and corrected for
target mass effects. The open circles in the left panel include the elastic contribution,
while the filled circles are only integrated over $W > 1.08$ GeV.
The right panel shows $\Gamma_1^{3He}$ and $\Gamma_1^n$ of the resonance region from Hall A E01-012, together with lower $Q^2$ results from E94-010, compared with 
the world DIS fits.
}
\label{fig:dualgam1}
\end{center}
\end{figure}

Detailed studies of duality in the spin structure functions $g_1^p$ and
$g_1^d$ have been published by the CLAS 
eg1~\cite{eg1dual} collaboration.
One observes a clear trend of strong, resonant deviations from the scaling 
curve at lower $Q^2$ towards a pretty good agreement at intermediate $Q^2$. 
The integral
of $g_1$ over the whole resonance region begins to agree with the NLO results 
above
$Q^2 \approx 1.7$ GeV$^2$ (see Fig.~\ref{fig:dualgam1}). 
The results 
on the proton and deuteron from eg1b~\cite{eg1dual} thus
indicate a much slower approach to ``global'' duality for the polarized structure
function $g_1$ than has been observed for the unpolarized structure 
functions~\cite{duality}
Local duality seems violated in the $\Delta-$resonance region even for $Q^2-$values as high as 5 GeV$^2$. 

The data taken
by the RSS collaboration in Hall C~\cite{RSS} 
corroborate these
observations and add more precise data points for $Q^2 \approx 1.3$ GeV$^2$.

The spin structure functions $g_1$ and $A_1$ were measured in the resonance
region ($W <2$ GeV) for $^3$He in the $Q^2-$region below 1 GeV$^2$ by
Hall A experiments E94-010~\cite{e94010-2,e94010-He3} and from 1 GeV$^2$ to 4 GeV$^2$ by 
E01-012~\cite{e02012-1}. 
Due to the prominent contributions from the $\Delta$ resonance, 
local duality does not appear to work at low $Q^2$ (below 2 GeV$^2$). 
At high $Q^2$ (above 2 GeV$^2$), the $\Delta$-resonance
contribution starts to diminish. 
The resonance data were integrated to study global duality. 
Figure~\ref{fig:dualgam1} (right panel) shows the results for both $^3$He and the neutron in comparison with the DIS fits evolved to the same $Q^2$. The resonance data agree with the DIS fits at least for $Q^2$ higher than 1.8 GeV$^2$, indicating that global duality holds for the neutron and $^3$He spin structure function, $g_1$, in the high $Q^2-$region (above 1.8 GeV$^2$).

The study of quark-hadron duality helps us to study higher-twist effects 
and decide where to extend the kinematic region 
apply partonic interpretations. 
From a practical point of view, the good understanding of the higher
twist effects and quark-hadron duality allows us to considerably 
extend the experimental database used to extract the polarized parton 
distributions~\cite{LSS}. 

\subsection{The effective strong coupling at large distance} \label{alpha_s}

\begin{figure}
\begin{center}
\includegraphics[scale=0.35]{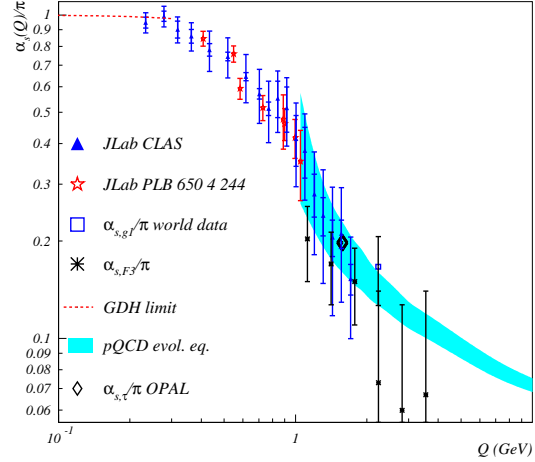}
\caption{\label{fig: alpha_s eff} 
Value of $\alpha_{s}^{eff}/\pi$ 
extracted from the data on the Bjorken sum, on $\tau$ decay and on the
Gross-Llewellyn-Smith sum.
The values of $\alpha_{s}^{eff}$ computed within pQCD and using the Bjorken sum 
are given by the gray band. The values of 
$\alpha_{s}^{eff}/\pi$ extracted using the GDH sum rule are given by the red dashed line.} 
\end{center}
\end{figure}

In QCD, the magnitude of the strong force is given by 
the running coupling constant $\alpha_{s}$. 
At large $Q^2$, in the pQCD domain, $\alpha_{s}$ is well defined and can be
experimentally extracted, e.g. using the Bjorken sum rule, 
(see eq. \ref{eq:genBj}).
The pQCD definition leads to an infinite coupling at large distances, 
where $Q^2$ approaches $\Lambda^{2}_{QCD}$, 
This is not a conceptual problem because we are out of the validity 
domain of pQCD. Since the data show no sign of discontinuity when crossing 
the intermediate $Q^{2}-$domain, see e.g. Fig.~\ref{fig:gamma1pn}, 
it is natural to look for a definition of an effective coupling
$\alpha_{s}^{eff}$, which works at any $Q^2$ and matches $\alpha_{s}$ 
at large $Q^{2}$ but stays finite at small $Q^{2}$.
The Bjorken sum rule can be used advantageously to define 
$\alpha_{s}^{eff}$ at 
low Q$^{2}$~\cite{alphaseff}. The data on the Bjorken sum are
used to experimentally extract $\alpha_{s}^{eff}$ following a prescription 
by Grunberg~\cite{grunberg}, see Fig.~\ref{fig: alpha_s eff}.
The Bjorken and GDH sum rules also allow us to determine $\alpha_{s}^{eff}$
at, respectively, large $Q^2$ and $Q^2 \simeq 0$. 
The extracted $\alpha_{s}^{eff}$ provides for the first time an experimental 
``effective coupling'' at all $Q^{2}$. An interesting feature 
is that $\alpha_s^{eff}$ becomes scale invariant at small $Q^{2}$, which 
was predicted by a number of calculations. 

\section{Conclusion}

A large body of nucleon spin-dependent cross-section and 
asymmetry data have been collected at low to moderate $Q^2$ in the resonance region. 
These data have been used to evaluate the $Q^2-$dependence of moments of 
the nucleon spin structure functions $g_1$ and $g_2$, including the 
GDH integral, the Bjorken sum, the BC sum and the spin polarizabilities, 
and to extract higher-twist contributions. The latter have 
provided access to the color polarizabilities in the nucleon. 
  
At $Q^2$ close to zero, available next-to-leading order $\chi$PT calculations
were tested against the data and found to be in reasonable agreement for $Q^2$ =
0.1 GeV$^2$ for the GDH integral $I(Q^2)$ and $\Gamma_1(Q^2)$. Above $Q^2$ =
0.1 GeV$^2$ a significant 
difference between the calculation and the data is observed, pointing to the
limit of applicability of $\chi$PT as $Q^2$ becomes larger. 
The comparison of results for the forward spin
polarizabilities $ \gamma_0(Q^2)$ and $\delta_{LT}$ with the 
$\chi$PT calculations show significant disagreements. $\gamma_0(Q^2)$
has significant contributions from the $\Delta$ resonance and its inclusion in
$\chi$PT calculations have not been fully worked out. The disagreements may 
indicate a better treatment of the $\Delta$ resonance is needed in the 
$\chi$PT calculations.
On the other hand, 
the $\chi$PT calculation of $\delta_{LT}$ was expected to offer a faster
convergence because of the absence of the $\Delta$ contribution. However, 
none of the available $\chi$PT
calculations can  reproduce $\delta_{LT}$ for the neutron 
at the lowest $Q^2$ point of 
0.1 GeV$^2$.  This discrepancy presents a significant challenge
to our theoretical understanding at its present level of approximations. 

Overall, the trend of the data  is well described by phenomenological 
models. The dramatic $Q^2-$evolution of $I_{GDH}$ from high to  low $Q^2$ was
observed as predicted by these models for both the proton and the neutron. This 
behavior is mainly determined by the relative strength and sign of the $\Delta$ 
resonance compared to that of higher-energy resonances and deep-inelastic processes.
This also shows that the current level of phenomenological understanding of the
resonance spin structure using these moments as observables is reasonable. 

The neutron BC sum rule is observed to be verified within experimental
uncertainties in the
low-$Q^2$ region due to a cancellation between the inelastic and the
elastic contributions. The BC sum rule is expected to be valid at all $Q^2$.
This test validates the assumptions going into the BC sum rule,
which provides confidence in sum rules with similar assumptions.

In the $Q^2$-region above 0.5 GeV$^2$, the 
first moment of $g_1$ for the proton, neutron and the proton-neutron 
difference were re-evaluated using the world data and the same extrapolation 
method of the unmeasured regions
for consistency. Then, in the framework of the OPE, the higher-twist $f_2$ and $d_2$ matrix 
elements were extracted. The low-$Q^2$ data have allowed us to gauge the convergence of the 
expansion used in this analysis. The extracted higher-twist (twist-4 and above) effects are not significant for $Q^2$ above 1 GeV$^2$. This fact may be related to the observation that quark-hadron duality works reasonably well 
at Q$^2$ above 1 GeV$^2$.

Finally, the proton and neutron electric and magnetic 
color polarizabilities were determined by combining the twist-4 matrix element $f_2$
and the twist-3 matrix element $d_2$ from the world data. Our findings show 
 a small and slightly positive value of $\chi_E$ and a value of $\chi_B$ close to zero 
for the neutron, while the proton has slightly larger values for both 
$\chi_E$ and $\chi_B$ but with opposite signs.

Overall, 
the new JLab data have provided valuable information on the transition between the non-perturbative 
to the perturbative regime of QCD. They form a precise data set for twist
expansion analysis and a check of $\chi$PT calculations.

Recently completed measurements~\cite{d2n,SANE} and future 12 GeV 
experiments~\cite{d2n} on the
$g_1$, $g_2$ structure functions and $d_2$ moments for the proton and 
neutron at $Q^2 \approx 1-5$~GeV$^2$
will reduce the uncertainty in the extracted higher-twist coefficients
and provide a benchmark test of Lattice QCD.

\section*{Acknowledgments}
Thanks to Alexandre Deur, Karl Slifer and Patricia Solvignon for providing 
figures. Thanks to Alexandre Deur, Kees de Jager and Vince Sulkosky for careful
proof-reading. This work was supported by the U.S. Department of Energy (DOE).
The Southeastern Universities
Research Association operates the Thomas Jefferson National Accelerator
Facility for the DOE under contract DE-AC05-84ER40150, Modification No. 175.



\begin{thebibliography}{0}
\bibitem{CPT} V. Bernard, Prog. Part. Nucl. Phys. {\bf 60}, 82 (2008).
\bibitem{LQCD} M. Gockeler {\em et al.}, PoS  {\bf LAT2006}, 120 (2006).
\bibitem{SD}I. C. Cloet and C. D. Roberts, PoS {\bf LC2008}, 047 (2008).
\bibitem{AdS} See, {\it e.g.}, J. Polchinski and M. J. Strassler, 
Phys. Rev. Lett. {\bf 88}, 031601 (2002); {\it ibid.} S. J. Brodsky and G. F. de Teramond, Phys. Rev. Lett. {\bf 96}, 201601 (2006); {\bf 94}, 201601 (2005). 
\bibitem{SV}E.~V.~Shuryak and A.~I.~Vainshtein,
Nucl. Phys. {\bf B 201}, 141 (1982). 
\bibitem{stein} E.~Stein, P.~Gornicki, L.~Mankiewicz and A.~Sch\"afer,
Phys. Lett. {\bf B 353}, 107 (1995).
\bibitem{Ji93} X. Ji, Nucl. Phys. {\bf B 402}, 217 (1993).
\bibitem{SG} O. Stern and W. Gerlach, Z. Physik {\bf 7}, 249 (1921); 
{\it ibid.}, {\bf 8}, 110 (1921); {\it ibid.}, {\bf 9}, 349 and 353 (1922).
 \bibitem{AMM} R. Frisch and O. Stern, Z. Physik {\bf 85}, 4 (1933);
I. Estermann and O. Stern, {\it ibid.}, 7 (1933).
\bibitem{GDH} S. B. Gerasimov, Sov. J. Nucl. Phys. {\bf 2}, 598 (1965)
\bibitem{GDH2} S. D. Drell and A. C. Hearn, Phys. Rev. Lett. {\bf 16}, 908 (1966).
\bibitem{Hof}R. Hofstadter, Rev. Mod. Phys. {\bf 28}, 214 (1956). 
\bibitem{KCL09} See, {\it e.g.}, S. E. Kuhn, J. -P. Chen and E. Leader, Prog. Part. Nucl. Phys., {\bf 63}, 1 (2009).
\bibitem{bjo66} J. D. Bjorken, Phys. Rev. {\bf 148}, 1467 (1966).
\bibitem{e97110} JLab E97-110, J. P. Chen, A. Deur, F. Garibaldi, 
spokespersons; V. Sulkosky, Proc. Spin Structure at Long Distance, Edited by 
J. P. Chen, W. Melnitchouk and K. Slifer, AIP {\bf 1151}, 93 (2009).
\bibitem{e94010-1} M. Amarian {\em et al.}, Phys. Rev. Lett. {\bf 89}, 242301 (2002).
\bibitem{e94010-2} M. Amarian {\em et al.}, Phys. Rev. Lett.
{\bf 92}, 022301 (2004).
 \bibitem{e94010-3} M. Amarian {\em et al.}, Phys. Rev. Lett. {\bf 93}, 152301 (2004).
 \bibitem{e94010-He3} K. Slifer {\em et al.}, Phys. Rev. Lett. {\bf 101}, 022303 (2008).
\bibitem{e02012-1} P. Solvignon {\em et al.}, Phys. Rev. Lett. {\bf 101}, 182502 (2008). 
\bibitem{e02012-2}P. Solvignon, Proc. Spin Structure at Long Distance, Edited by 
J. P. Chen, W. Melnitchouk and K. Slifer, AIP {\bf 1151}, 101 (2009).
\bibitem{eg1p} R. Fatemi {\em et al.}, Phys. Rev. Lett. {\bf 91}, 222002 (2003).
\bibitem{eg1d} J. Yun {\em et al.}, Phys. Rev. {\bf C 67}, 055204 (2003). 
\bibitem{eg1b} K. V. Dharmawardane {\em et al.}, Phys. Lett. {\bf B 641}, 11 (2006).
\bibitem{eg1gam0} Y. Prok {\em et al.}, Phys. Lett. {\bf B 672}, 12 (2009). 
 \bibitem{eg1dual} P. Bosted {\em et al.}, Phys. Rev. {\bf C 75}, 035203 (2007). 
\bibitem{RSS} F. R. Wesselmann {\em et al.}, Phys. Rev. Lett. {\bf 98}, 132003 (2007).
\bibitem{RSS-2} K. Slifer {\em et al.}, arXiv nucl-ex/08120031. 
\bibitem{HTN} Z.-E. Meziani {\em et al.}, Phys. Lett. {\bf B 613}, 148 (2005).
\bibitem{HTP} M. Osipenko {\em et al.}, Phys. Lett. {\bf B 609}, 259 (2005).
\bibitem{HTP2} A. Deur, arXiv nucl-ex/0508022.
\bibitem{BJ} A. Deur {\em et al.}, Phys. Rev. Lett. {\bf 93}, 212001 (2004);
A. Deur {\em et al.}, Phys. Rev. {\bf D 78}, 032001 (2008).
\bibitem{ahr01} J. Ahrens {\it et al.} Phys. Rev. Lett. {\bf 87}, 022003 (2001).\bibitem{dutz03}
 H. Dutz {\it et al.}, Phys. Rev. Lett. {\bf 91}, 192001 (2003).
\bibitem{dutz04}
 H. Dutz {\it et al.}, Phys. Rev. Lett. {\bf 93}, 032003 (2004).
\bibitem{Anselmino:1994gn} M. Anselmino, A. Efremov and E. Leader, 
Phys. Rep. {\bf 261}, 1 (1995); {\it ibid.}, {\bf 281} 399 (erratum) (1997).  
\bibitem{Soffer:1999zv} J. Soffer, O. V. Teryaev, arXiv hep-ph/9906455.
\bibitem{Artru:2008cp}
 X.~Artru, M.~Elchikh, J.~M.~Richard, J.~Soffer and O.~V.~Teryaev,
  hep-ph/08020164.
\bibitem{DW08} D. Drechsel and T. Walcher, Rev. Mod. Phys. {\bf 80}, 731 (2008). 
\bibitem{hand} L. N. Hand, Phys. Rev. {\bf 129}, 1834 (1963).
\bibitem{Hoodbhoy:1988am} P. Hoodbhoy, R. L. Jaffe and A. Manohar, Nucl. Phys. 
{\bf B 312}, 571 (1989).
\bibitem{Frankfurt:1983qs} L. L. Frankfurt and M. I. Strikman, Nucl. Phys. 
{\bf A 405}, 557 (1983).  
\bibitem{dre01} D. Drechsel, S.S. Kamalov and L. Tiator, Phys. Rev. {\bf D 63}, 114010 (2001). 
\bibitem{ji01} X. Ji and J. Osborne, J. of Phys. G
{\bf 27}, 127 (2001).
\bibitem{ans89} M. Anselmino, B. L. Ioffe, and E. Leader, Sov. J. Nucl. Phys. {\bf 49}, 136 (1989).
\bibitem{dre03} D. Drechsel, B. Pasquini and M. Vanderhaeghen, Phys. Rep. 
{\bf 378}, 99 (2003).
\bibitem{dre04} D. Drechsel and L. Tiator, Ann. Rev. Nucl. Part. Sci.
{\bf 54}, 69 (2004).
\bibitem{BD} J. D. Bjorken and S. D. Drell, ``Relativistic Quantum Fields'',
McGraw Hill, New York (1965). 
\bibitem{low} F. E. Low, Phys. Rev. {\bf 96}, 1428 (1954).
\bibitem{ope} K. Wilson, Phys. Rev. {\bf 179}, 1499 (1969).
\bibitem{BC} H. Burkhardt and W. N. Cottingham, Ann. Phys. (N.Y.) {\bf 56}, 453 (1970).
\bibitem{EJ} J. Ellis and R. L. Jaffe, Phys. Rev. {\bf D 9}, 1444 (1974); 
{\it ibid.}, {\bf D 10}, 1669 (1974).
\bibitem{JU} X. Ji and P. Unrau, Phys. Lett. {\bf B 333}, 228 (1994). 
\bibitem{JM} X. Ji and W. Melnitchouk, Phys. Rev. {\bf D 56},1 (1997).
\bibitem{WW} S. Wandzura and F. Wilczek, Phys. Lett.  {\bf B 72}, 195 (1977).
\bibitem{burkardt} M. Burkardt, Proc. Spin Structure at Long Distance, Edited by 
J. P. Chen, W. Melnitchouk and K. Slifer, AIP {\bf 1151}, 26 (2009). 
\bibitem{EMC} J. Ashman {\em et al.}, Phys. Lett. {\bf B 206}, 364 (1988). 
\bibitem{SMCp} SMC collaboration: D. Adams \emph{et al.}, Phys. Lett. {\bf B 329}, 399 (1994).
\bibitem{SMCd} SMC collaboration: D. Adams \emph{et al.}, Phys. Lett. {\bf B 357}, 248 (1995).
\bibitem{SMCd2} SMC collaboration: D. Adams \emph{et al.}, Phys. Lett.{\bf B 396}, 338 (1997).
\bibitem{SMCd0} SMC collaboration: D. Adeva \emph{et al.}, Phys. Lett. {\bf B 302}, 533 (1993).
\bibitem{SMCp2} SMC collaboration: D. Adeva \emph{et al.}, Phys. Lett.{\bf B 412}, 414 (1997).
\bibitem{HERMESn} K. Ackerstaff {\em et al.}, Phys. Lett. {\bf B 404}, 383 (1997).
\bibitem{HERMESGDH} K. Ackerstaff {\em et al.}, Phys. Lett. {\bf B 444}, 531 (1998).
\bibitem{HERMESp} A. Airapetian, {\em et al.}, Phys. Lett. {\bf B 442}, 484 (1998).
\bibitem{E80} M. J. Alguard {\em et al.}, Phys. Rev. Lett. {\bf 37}, 1261 (1976).
\bibitem{E130} M. J. Alguard {\em et al.}, Phys. Rev. Lett. {\bf 41}, 70 (1976).
\bibitem{E142} P. L. Anthony {\em et al.}, Phys. Rev. {\bf D 54}, 6620 (1996).
\bibitem{E143} K. Abe {\em et al.},
Phys. Rev. {\bf D 58},112003 (1998). 
\bibitem{E154} K. Abe {\em et al.},
Phys. Rev. Lett. {\bf 79}, 26 (1997).
\bibitem{E154g2} K. Abe {\em et al.}, Phys. Lett. {\bf B 404}, 377 (1997).
\bibitem{E155} P. L. Anthony, {\em et al.}, Phys. Lett. {\bf B 493}, 19 (2000).
\bibitem{E155g2} P. L. Anthony, {\em et al.}, Phys. Lett. {\bf B 458}, 529 (1999).
\bibitem{E155x} P. L. Anthony {\em et al.}, Phys. Lett. {\bf B 553}, 18 (2003).
\bibitem{COMPASS} C. Bernet, Proceedings of DIS2005 (2005), Madison, Wisconsin,
AIP Conf. Proc. (2005).
\bibitem{RHICspin} A. Deshpande, Proceedings of DIS2005 (2005), Madison, Wisconsin, AIP Conf. Proc. (2005).
\bibitem{ji97} X. Ji, Phys. Rev. Lett. {\bf 78}, 610 (1997).
\bibitem{Bur05}M. Burkardt, Phys. Rev. {\bf D 72}, 094020 (2005).
\bibitem{HallA nim}
Hall A collaboration: J. Alcorn \emph{et al.}, 
Nucl. Inst. Meth. \textbf{A522}, 294 (2004).
\bibitem{Astatus07} Hall A Status Report - 2007.
\bibitem{HallB nim}
CLAS collaboration: B. A. Mecking \emph{et al.},
Nucl. Inst. Meth. \textbf{A503}, 513 (2003).
\bibitem{mo69} L.W. Mo and Y.S. Tsai, Rev. Mod. Phys. {\bf 41}, 205 (1969).
\bibitem{aku94} I.V Akushevich and N.M. Shumeiko, J. Phys.  {\bf G 20}, 513 (1994).
\bibitem{cio97} C. Ciofi degli Atti and S. Scopetta, Phys. Lett. {\bf B 404}, 223 (1997). 
\bibitem{DNP} 
A. Abragam and M. Goldman, Rep. Prog. Phys. \textbf{41} 396 (1978).
\bibitem{crabb}
D.G. Crabb and W. Meyer, Annu. Rev. Nucl. Part. Sci. \textbf{47}, 67 (1997).
\bibitem{NH3} 
C. D. Keith \emph{et al}., Nucl. Inst. Meth. A \textbf{501}, 327 (2003).
\bibitem{F2&RSLAC} L.W. Whitlow {\it et al.}, Phys. Lett. {\bf B 250}, 193 (1990);
\bibitem{F2&RJLab}
Y. Liang et al., nucl-ex/0410027.
\bibitem{RCSLACPOL} 
K. Abe \emph{et al}., Phys. Rev. \textbf{D 58} 112003 (1998).
\bibitem{NMC} M. Arneodo \emph{et al}.,  Phys. Lett. {\bf B 364}, 107 (1995).
\bibitem{AO} V. Burkert and Z. Li, Phys. Rev \textbf{D 47}, 46 (1993).
\bibitem{Sofferbound} J. Soffer, Phys. Rev. Lett. \textbf{74}, 1292 (1995).
\bibitem{tho00} N. Bianchi and  E. Thomas, Nucl. Phys. {\bf B 82} (Proc. Suppl.), 256 (2000).
\bibitem{ji00} X. Ji, C. Kao, and J. Osborne, Phys. Lett. B {\bf 472}, 1 (2000).\bibitem{ber02} V. Bernard, T. Hemmert and Ulf-G. Meissner, Phys. Lett. 
{\bf B 545}, 105 (2002)
\bibitem{ber03} V. Bernard, T. Hemmert and Ulf-G. Meissner,
Phys. Rev. {\bf D 67}, 076008 (2003).
\bibitem{krebs09}H. Krebs, {\em et al.}, Proc. Spin Structure at Long Distance, Edited by 
J. P. Chen, W. Melnitchouk and K. Slifer, AIP {\bf 1151}, 42 (2009).
 \bibitem{kao08} C. W. Kao, Proc. 18th Int. Spin Phys. Symp. Edited by D. Crabb, {\em et al.}, AIP {\bf 1149}, 289 (2008). 
\bibitem{sof02}J. Soffer and O. V. Teryaev, Phys. Rev. {\bf D 70}, 116004 (2004).
\bibitem{bur92}V. D. Burkert and B. L. Ioffe, Phys. Lett. {\bf B 296}, 223 (1992).
\bibitem{12gev}
The Science Driving the 12 GeV Upgrade of CEBAF, \\
http://www.jlab.org/div\_dept/physics\_division/GeV.html.
\bibitem{mer96} P. Mergell, Ulf-G. Meissner and D. Drechsel, Nucl. Phys.  {\bf A 596}, 367 (1996).
\bibitem{mel03} S. A. Kulagin and W. Melnitchouk (to be published), private communication. 
\bibitem{van02} C. W. Kao, T. Spitzenberg and M. Vanderhaeghen, Phys. Rev. {\bf D  67}, 016001 (2003).
\bibitem{goc01} M. Gockeler {\em et al.}, Phys. Rev. {\bf D 63}, 074506, (2001).
\bibitem{bur01}
V. D. Burkert,
Phys. Rev. \textbf{D 63}, 097904 (2001).
\bibitem{BB} J. Blumlein and H. Boettcher, Nucl. Phys. \textbf{636}, 225 (2002).
\bibitem{e97103} K. Kramer {\em et al.}, Phys. Rev. Lett. {\bf 95}, 142002 (2005).
\bibitem{Zheng} X. Zheng {\em et al.}, Phys. Rev. {\bf C 70}, 065207 (2004). 
\bibitem{duality} W. Melnitchouk, R. Ent and C. Keppel, Phys. Rept. {\bf 406},
127 (2005).
\bibitem{LSS} E. Leader, A. V. Sidorov, D. B. Stamenov, Phys. Rev. {\bf D 63},
034023 (2006). 
\bibitem{alphaseff} A. Deur, V. Burkert, J. P. Chen and W. Korsch, Phys. Lett.
{\bf B 665}, 349 (2008);{\bf B 650}, 244 (2007).
 \bibitem{grunberg} G. Grunberg, Phys. Lett. {\bf B 95}, 70 (1980); 
 Phys. Rev. {\bf D 29}, 2315 (1984);
Phys. Rev. {\bf D 40}, 680 (1989).
\bibitem{d2n} B. Sawatzky {\em et al.}, Proc. Spin Structure at Long Distance, Edited by 
J. P. Chen, W. Melnitchouk and K. Slifer, AIP {\bf 1151}, 145 (2009). 
\bibitem{SANE} O. A. Randon, Proc. Spin Structure at Long Distance, Edited by 
J. P. Chen, W. Melnitchouk and K. Slifer, AIP {\bf 1151}, 82 (2009).
\end{thebibliography}
\end{document}